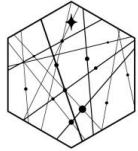

# Outer Space Cyberattacks

Generating Novel Scenarios to Avoid Surprise

A report of the Ethics + Emerging Sciences Group

Authors:         Patrick Lin
                 Keith Abney
                 Bruce DeBruhl
                 Kira Abercromby
                 Henry Danielson
                 Ryan Jenkins

Project site:    https://**space**cybersecurity.org/

Draft date:      17 June 2024

Version:         1.0.0

**Ethics + Emerging Sciences Group** | CAL POLY

# Table of Contents

# Author biographies

**Patrick Lin**, PhD, is the director of the Ethics + Emerging Sciences Group at Cal Poly, where he is a philosophy professor. He also serves on the US National Space Council's Users' Advisory Group (UAG) and is affiliated with Stanford Law School, Czech Academy of Sciences, World Economic Forum, Aurelia Institute, For All Moonkind, AIAA, and others. Previous affiliations include Stanford Engineering, US Naval Academy, Dartmouth, Univ. of Iceland (Fulbright specialist), Center for a New American Security, New America Foundation, UNIDIR, 100 Year Study on AI, and more. Prof. Lin is well published in technology ethics, incl. cyber, space development, AI, robotics, military systems, bioengineering, materials science, and more.

**Keith Abney**, ABD, is a senior fellow of the Ethics + Emerging Sciences Group at Cal Poly, San Luis Obispo, and senior lecturer in the Philosophy Department; he both teaches and publishes on technology and biomedical ethics, including space security and war, space colonization, and settlement, human enhancement, AI, robotics, and more. He co-edited a special journal edition on "Human Colonization of Other Worlds", as well as the books *Robot Ethics* (MIT Press, 2012) and *Robot Ethics 2.0* (Oxford University Press, 2017), and has co-authored funded reports on autonomous military robotics, military human enhancements, AI ethics, and cyberattacks. He also serves as a member of the bioethics committee of Arroyo Grande Community Hospital.

**Bruce DeBruhl**, PhD, is an associate professor at Cal Poly, San Luis Obispo, in the Computer Science and Software Engineering Department, as well as the Computer Engineering Department. He is also an advanced computer scientist at SRI International. His educational goal is to develop opportunities for diverse students to get hands on experience with security and privacy. This has included teaching major and non-major courses in privacy engineering, network security, software security, binary exploitation, hardware security, and cybersecurity policy. Dr. DeBruhl's research interests include wireless security, cyber-physical security, location privacy, automotive security, and malware for cyber-physical systems. He received his PhD and MS degrees in Electrical and Computer Engineering at Carnegie Mellon University Silicon Valley and his BS degree in Electrical Engineering from Kettering University.

**Kira Jorgensen Abercromby**, PhD, is a professor at Cal Poly, San Luis Obispo, in the Aerospace Engineering Department. She has taught multiple classes in orbital mechanics, spacecraft controls, and spacecraft environmental effects on spacecraft. Prior to joining Cal Poly, Dr. Abercromby worked at NASA Johnson Space Center in the Orbital Debris Program Office in Houston, Texas. Her research interests include human-made space debris, orbit determination, and space environmental effects on spacecraft. She received her PhD and MS degrees in Aerospace Engineering at University of Colorado, Boulder, and BS degree in Astrophysics from UCLA.

**Henry Danielson**, MS, has a broad depth of knowledge in cybersecurity/computer security and has obtained his Certified Information Systems Security Officer (CISSO). His current roles include serving as a technical advisor at the California Cybersecurity Institute (CCI), as well as a lecturer at Cal Poly, San Luis Obispo. Mr. Danielson is also a Goon at DEFCON, the largest hacking conference in the world. He is part of the Aerospace Village at DEFCON and works with Vint Cerf on the Interplanetary Networking Special Interest Group (IPNSIG) Academy working group and the Technical Documentation working group. He focuses on developing cybersecurity curriculum and helping manage CCI's annual Space Grand Challenge (gamified capture-the-flag), and he has taught many courses on information security. Mr. Danielson has collaborated with the United States Space Force, Jet Propulsion Laboratories (JPL), and Space Systems Command to help develop Space Grand Challenge.

**Ryan Jenkins,** PhD, is a full professor, associate chair of philosophy, and associate director of the Ethics + Emerging Sciences Group at Cal Poly. His research focuses on the potential for emerging technologies to enable or encumber meaningful human lives—especially artificial intelligence and robotics. Dr. Jenkins is a former member of the IEEE TechEthics Ad Hoc committee and a former co-chair of the Robot Ethics Technical Committee of the IEEE's Robotics & Automation Society. He has served as principal investigator or senior personnel for several grants on the ethics of autonomous vehicles, predictive policing, cyberwar, AI kitchens/robot cooks, and more. His work has appeared in journals such as *Techné*, *Ethical Theory and Moral Practice*, and the *Journal of Military Ethics*, as well as public fora including the *Washington Post*, *Slate*, and *Forbes*.

.....................

# About the Ethics + Emerging Sciences Group

Established in 2007 at California Polytechnic State University (Cal Poly), San Luis Obispo, the Ethics + Emerging Sciences Group is a non-partisan think tank focused on risk, ethics, and social concerns related to new sciences and technologies. This includes AI and robotics, bioengineering, cybersecurity, digital media, materials science, and more, especially in security and defense applications. Our work in "frontier ethics" aims to help develop norms and governance in the Arctic, outer space, and other domains, as emerging technologies can unlock greater access to those frontiers. Site: https://ethics.calpoly.edu/

# Acknowledgements

This material is based upon work supported by the US National Science Foundation under grant no. 2208458.

We thank our research assistants at Cal Poly: Scott Collier and Sage Meadows. And we thank the space experts at Project Lodestar for their creative contributions: Alires Almon, Jason Batt, and Jaym Gates.

Our work benefitted from discussions with the many distinguished researchers at our [expert workshop](#) at Stanford University, 30-31 March 2023, particularly Herb Lin and Corie Wieland as our hosts, with material support from Stanford CISAC and the Hoover Institution.

We are indebted to James Pavur and Sara Langston for their invaluable input throughout this report. And we are grateful for the discussions and feedback of other experts along the way, including: Timiebi Aganaba, Gil Baram, Joe Bassi, Martin Minnich, C. Tony Pfaff, Nick Reese, as well as personnel at NASA Headquarters, NASA Jet Propulsion Laboratory, Vandenberg Space Force Base, and the National Space Council's Users' Advisory Group.

Finally, we acknowledge Cal Poly, San Luis Obispo, particularly the Philosophy Department (College of Liberal Arts), Computer Science Department (College of Engineering), and the California Cybersecurity Institute for their continuing support, both moral and material.

Any opinions, findings, and conclusions or recommendations expressed in this material are those of the authors and do not necessarily reflect the views of the US National Science Foundation or any other organization or person mentioned here.

# Acronyms list

| | |
|---|---|
| **2FA** | Two-Factor Authentication |
| **AI** | Artificial Intelligence |
| **API** | Application Programming Interface |
| **APT** | Advanced Persistent Threats |
| **ASAT** | Anti-Satellite |
| **C3I** | Command, Communications, Control, and Intelligence |
| **CFAA** | Computer Fraud and Abuse Act |
| **CME** | Coronal Mass Ejection |
| **COTS** | Commercial Off-The-Shelf |
| **DDoS** | Distributed Denial-of-Service |
| **ESA** | European Space Agency |
| **ET** | Extraterrestrial |
| **GEO** | Geosynchronous Orbit |
| **GNSS** | Global Navigation Satellite System |
| **GOES** | Geostationary Operational Environmental Satellite |
| **GPS** | Global Positioning Systems |
| **IoT** | Internet of Things |
| **IP** | Internet Protocol |
| **ISP** | Internet Service Provider |
| **ISS** | International Space Station |
| **IT** | Information Technology |
| **ITU** | International Telecommunications Union |
| **JWST** | James Webb Space Telescope |
| **LEO** | Low Earth Orbit |
| **LIDAR** | Light Detection and Ranging |
| **LLM** | Large Language Model |
| **LSP** | Launch Service Provider |
| **MEO** | Medium Earth Orbit |
| **METI** | Messaging Extraterrestrial Intelligence |
| **MITM** | Man-in-the-Middle |
| **ML** | Machine Learning |
| **NASA** | National Aeronautics and Space Administration |
| **NRO** | National Reconnaissance Office |
| **OSINT** | Open-Source Intelligence |
| **OST** | Outer Space Treaty |
| **OT** | Operational Technology |
| **PMC** | Private Military Contractor |
| **PNT** | Position, Navigation, and Timing |
| **PPI** | Personal Identifiable Information |
| **R&D** | Research and Development |
| **RPO** | Rendezvous and Proximity Operations |
| **SBSP** | Space-Based Solar Power |
| **SETI** | Search for Extraterrestrial Intelligence |
| **SOE** | State-Owned Entities |
| **SSN** | Space Surveillance Network |
| **STM** | Space Traffic Management |
| **STS** | Science and Technology Studies |
| **SWOT** | Surface Water and Ocean Topography |
| **UAP** | Unidentified Aerial Phenomena |
| **VPN** | Virtual Private Network |

# Executive summary

Though general awareness around it may be low, space cyberattacks are an increasingly urgent problem given the vital role that space systems play in the modern world.  Open-source or public discussions about it typically revolve around only a couple generic scenarios, namely satellite hacking and signals jamming or spoofing.  But there are *so many more* possibilities.

The report offers a scenario-prompt generator—a taxonomy of sorts, called the **ICARUS matrix**—that can create more than 4 million unique scenario-prompts.  We will offer a starting set of 42 scenarios, briefly describing each one, to begin priming the imagination-pump so that many more researchers can bring their diverse expertise and perspectives to bear on the problem.

A failure to imagine novel scenarios is a major risk in being taken by surprise and severely harmed by threat actors who are constantly devising new ways, inventive and resourceful ways, to breach the digital systems that control our wired world.  To stay vigilant, defenders likewise need to be imaginative to keep up in this adversarial dance between hunter and prey in cybersecurity.

More than offering novel scenarios, we will also explore the drivers of the space cybersecurity problem, which include at least seven factors we have identified.  For instance, the shared threat of space debris would seem to push rational states and actors to avoid *kinetic* conflicts in orbit, which weighs in favor of cyberoperations as the dominant form of space conflicts.

Outer space is the next frontier for cybersecurity.  To guard against space cyberattacks, we need to understand and anticipate them, and imagination is at the very heart of both cybersecurity and frontiers.

> *I've never been certain whether the moral of the Icarus story should only be, as is generally accepted, "Don't try to fly too high", or whether it might also be thought of as, "Forget the wax and feathers, and do a better job on the wings."*
>
> – *Stanley Kubrick*[1]

# 01

# Introduction

A s inhospitable as outer space is to human life, we owe the modern world to it. Yet space is invisible—out of sight, out of mind—to most of us. After all, most people haven't been to outer space, and many people can go all day without even looking up at the sky. And space science might not be enough to inspire interest when *basic* science and facts about the world seem so hard for many to grasp today.

So, it's not a surprise that outer space *cybersecurity* hasn't gained much attention, though public awareness is rising.[2,3] Most people don't understand or even like to think about "regular" cybersecurity, which can also be as esoteric and intimidating as space science. If it's tempting to just ignore cybersecurity on Earth, even though our wired lives depend on it, then it's even easier to ignore cybersecurity in the faraway and alien domain of outer space.

But that would be a **strategic, dangerous mistake** given how much our world depends on space-enabled capabilities, again largely invisible to most of us.[4] Just checking the time on our smartphones relies on global positioning systems (GPS) to synchronize time worldwide. Less obviously, this satellite-enabled timing is crucial in financial services—from credit card transactions to stock exchanges—where every detail, such as time of payment or withdrawal, needs to be precisely captured and coordinated; money is literally at stake. Similarly, making a mobile phone call relies on this precise coordination of time, given how mobile communications networks work.

Of course, GPS is better known for its daily role in positioning and navigation for airplanes, boats, trucks, cars, people, and others. For instance, GPS is crucial for managing and coordinating fleets of trucks that transport goods to stock local stores every day. Satellites also are our "eyes in the skies" and beam back Earth-observation data that help us to forecast the weather, monitor environmental changes and animal populations, track and respond to natural disasters, boost agricultural crop yields, manage land and water use, surveil adversaries and troop movements, and so much more.

As many critical services and infrastructure are based in space, it's not an exaggeration to say that the modern world wouldn't exist without space capabilities, the loss of which could be fatal to vulnerable people. Economic and national security would be at serious risk, among other things we care about.

Fortunately, good people are thinking about space cybersecurity to defend our space assets from bad actors, both from technical and policy directions. But in these discussions, at least the *unclassified* ones, only a few generic scenarios are typically trotted out, which usually involve some vague hacking of a satellite or jamming/spoofing of signals, such as critical GPS communications.[†] As technological advancements are opening up access to space for more people, including private individuals, many more cybersecurity scenarios now seem plausible and need to be considered in this planning.

If we've learned anything in the short history of cybersecurity, it's that the cat-and-mouse game between attackers and defenders is constantly shifting; new, inventive exploits continue to surprise us over time. This report, therefore, serves to inform technical and policy planning in outer space cybersecurity by helping to **anticipate those surprises**. Without a fuller range of scenarios to consider and an organizing framework—which this report will offer—technical and policy solutions could be either too narrow (and less effective) or overly broad (and not nuanced enough) if they're limited to or misdirected at only a few obvious scenarios.

Undoubtedly, there will be many more possible scenarios. Though we will offer more than 40 novel scenarios, they are not meant to be a complete set, which isn't possible anyway. But the hope is that this discussion will be an "imagination-pump" to help conceive of other new scenarios, which in turn shines a brighter spotlight to attract more experts who can help address the looming space-cybersecurity problem.

Again, it's possible that *classified* conversations exist about novel space cyberattack scenarios, but that's unknown to the broader public. Security classification is often a key barrier toward a more expansive, probing discussion with a wider range of experts. This project seeks to be a springboard for that larger, non-classified conversation and analysis, so that both classified and non-classified planning can be more thoughtful and effective.

In the sections that follow, we will start by discussing **seven factors** that are driving the space-cybersecurity problem, which also explain why this is an urgent domain to attend to. Then we will offer a scenario-prompt generator for space cybersecurity—a taxonomy called the **ICARUS matrix**, an acronym for "**I**magining **C**yberattacks to **A**nticipate **R**isks **U**nique to **S**pace"—as well as brief descriptions of novel cyberattack scenarios and some critical-thinking questions to further develop and interrogate the scenarios.

---

[†] See our remarks on terminology in this section below, specifically on why we are including electromagnetic spectrum as relevant scenarios in cybersecurity.

In the rest of this section, we will clarify what we mean by "cybersecurity" and other key terms in this report.

## Terminology

For the purposes of this report, we will use "cyberattacks" to refer to unauthorized, **non-kinetic** or non-physical attacks that *involve* computing devices or systems, *either* as means or targets.  This is broader than the typical understanding in which cyberattacks must *use* computing devices as unauthorized *means* to access a system.  And by "attacks", we mean the usual understanding in cybersecurity in which the unauthorized intrusion aims to disable, disrupt, degrade, or otherwise compromise a computing system, including the confidentiality, integrity, or availability of its data, even if nothing else is corrupted or harmed.

As examples:  A solely *physical* attack on a computing system—whether by water, hammer, bullet, bomb, and so on—still wouldn't qualify as a cyberattack since it's *not* a non-physical attack (by definition).  But hacking a water system inside a spacecraft to spray water that destroys a digital control system would be a cyberattack, since a critical part of that attack-chain involves a cyberattack (the initial hacking).  Likewise, social engineering to trick users to reveal their passwords—with which hackers can gain unauthorized access to a computer network and steal its data or do worse—is still a part of a cyberattack since the attack on the information system itself would be non-physical, even if the necessary social-engineering tactic might depend on physical things, such as a planted USB stick that auto-runs spyware when inserted.

Given the above working definition, we will use "cybersecurity" to also include security related to **electromagnetic spectrum**, which is often treated as a separate but related domain in other discussions.  The rationale is to streamline our discussion, instead of using the more cumbersome "cybersecurity and spectrum security."  Major aerospace organizations, for instance, are also [adopting this position](#), in part to ensure that attention to spectrum security isn't lost for lack of a clear disciplinary home.[5]  Further, the difference between two is becoming more [blurry](#), and it isn't very important for the purposes of this report anyway, even if it's useful in other discussions.[6]

To wit, we recognize that signals jamming and spoofing aren't strictly "cyberattacks" but are more properly interference with radio signals or the electromagnetic spectrum.  In the context of armed conflicts, they wouldn't be part of cyberwarfare but fall in the domain of "electronic warfare."  However, since the twin domains are closely related enough in that they are both about non-physical attacks, and since radios can also be *software-defined* and therefore vulnerable to cyberattacks, not just spectrum interference, we will include signals jamming and spoofing as examples of cyberattacks, especially as they are important incidents or scenarios that have already occurred.

We won't discuss the basics of cybersecurity further here, including its history and methods, as those discussions are plentiful elsewhere and out of scope for this report. But this report does assume some basic familiarity with cybersecurity.

By "space cyberattacks", we don't necessarily mean cyberattacks on computer systems that reside in outer space, though that may be the first image that comes to mind. Instead, we will use this phrase to mean cyberattacks on *any part* of a space ecosystem, which may include:

- **Launch** segments, e.g., launch complexes, launch vehicles, payloads, R&D facilities;
- **Ground** segments, i.e., Earth-based infrastructure and services to support the functioning of a space system, such as mission control and personnel terminals;
- **Space** segments, i.e., space objects, such as satellites, space stations, telescopes;
- **User** segments, e.g., GPS receivers, satellite-internet terminals, satellite phones;
- **Link** segments, e.g., communications links from ground-to-space (either uplink to or downlink from a satellite), space-to-space (cross-link), or even ground-to-ground.

The following image illustrates these various segments in a space ecosystem:

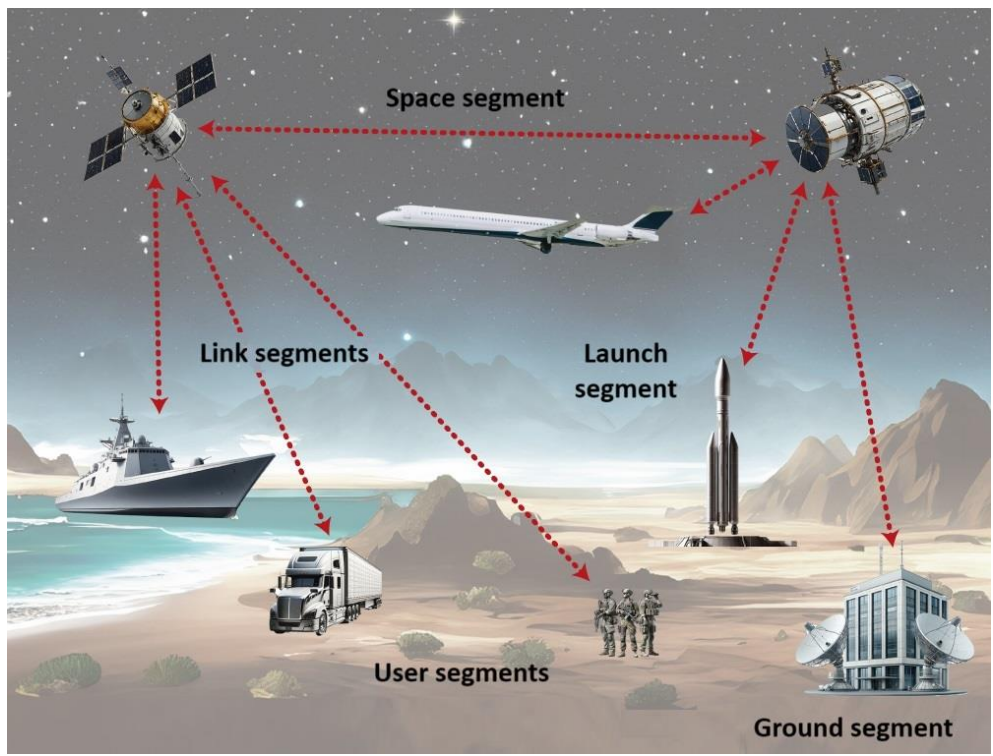

*Image 1: space system segments. Credit: Ethics + Emerging Sciences Group.*

For instance, the US Space Surveillance Network (SSN) detects and tracks artificial objects orbiting Earth, including space debris, returning spacecraft and deorbiting satellites, and possible missiles.[7]  This is vital for, say, space traffic management (STM), e.g., to anticipate collision courses with space debris and make emergency maneuvers for spacecrafts to avoid them.  The ground-based radars and optical telescopes are part of SSN's ground segment, while the satellites are part of its space segment.  The link segments could be any of the above: both to and from space, from satellite to satellite, or from ground station to ground station.

As another example, about an hour before Russia invaded Ukraine on 24 February 2022, Russia had reportedly conducted a cyberattack on Viasat, a US communications company, to cut off its satellite internet connectivity in Ukraine.[8]  This was an attack on Viasat's modems and routers, which means it was an attack on the *user segment* to disrupt the *link segments*, both uplink and downlink.  Though no space-based assets were directly targeted by this cyberattack, it still counts as a space cyberattack because the target was part of Viasat's space ecosystem.

That said, this report will predominantly focus on cybersecurity in **space segments**, since space assets are much harder to secure than computer systems and devices on the ground, as explained in the next section.  Moreover, those scenarios haven't been considered nearly as much as cybersecurity scenarios on Earth have been, which may be more familiar and ordinary.  For instance, a cyberattack on a space R&D facility on Earth may resemble a cyberattack on many other organizations and therefore would already be guarded against specifically.

Further, it's possible that not all cyberattacks on space organizations would count as a "space cyberattack" instead of a more usual kind.  An example would be if "space" were only incidental to a cyberattack that could just as well have been directed at a non-space company.  For instance, such an intrusion might aim to steal the social security numbers of employees for *ordinary* identity theft—perhaps not even knowing or caring that the employees worked at a space company—as opposed a multi-step attack meant to penetrate a space organization's IT systems to specifically compromise space data, a product, or a mission.  Precisely drawing this line is not important to this report, as we will be dealing with unambiguous instances of space cyberattacks.

Finally, by "satellite", we will refer to human-made or artificial space objects in an orbit.  If we are referring to a *natural* satellite, we will be explicit in that, e.g., to specify that a space object is an asteroid.

## 02

# Context: what is driving the problem?

The history of space cyberattacks dates back only a few decades, even though it's been nearly 70 years since the launch of Earth's first artificial satellite—the former Soviet Union's Sputnik 1 in 1957.[9]  This history followed the rise of personal computers and the internet in the 1980s and 1990s, which also saw the rapid emergence and proliferation of digital cyberattacks.  It was only a matter of time before computers in space would be targeted.

The world's first space cyberattack is commonly placed at 1986 and attributed to a "video pirate", a satellite operations engineer with the alias of Captain Midnight.[10]  His mission, which now seems quaint, was to protest the rising costs of HBO's satellite TV services by jamming HBO's transmission and broadcasting his own message for 4.5 minutes, as a modern-day Robin Hood figure.

Other early accounts of satellite hacking have been disputed.  Part of the problem is that it's not easy to detect a cyberattack in the first place, much less to attribute an attack to a particular actor, organization, or even nation given the lack of clear, physical evidence in cyberspace.  For instance, the following accounts have been disputed as having natural or non-cyber causes: in 1998, the German x-ray telescope satellite (ROSAT) was allegedly hacked to point its solar panels at the sun, which led to overcharging and failure of the batteries; and in 1999, British military communications satellites (Skynet) were allegedly hacked and held for ransom.[11,12]

But other cases are clearer, even if denials can be expected as a matter of course.  In 1999, a 15-year old hacker stole the source code from NASA servers that controlled the physical environment of the International Space Station (ISS); this cyberattack was undisputed and prosecuted.[13]  In 2006 (and again in 2011), UAE's Thuraya satellites were jammed to disrupt its satellite phone service, reportedly by Libya.[14,15]  In 2007 and 2008, NASA satellites were hacked, taking over control of Landsat-7 and Terra AM-1 satellites for several minutes; the cyberattack was attributed to China, though this was still also disputed.[16,17,18]

Fast-forward a few years, starting in 2011, Iran allegedly demonstrated its ability to spoof GPS signals to capture a US drone and disrupt civilian avionics.[19,20]  In 2014, the Palestinian group [Hamas](#) claimed responsibility in hacking an Israeli satellite-television transmission to broadcast its own message.[21]  In 2015, Russian hackers ([Turla](#)) were blamed for hijacking satellite communications; this is a notable, rare example of attacking space systems in order to attack traditional computers on the ground, by cloaking its data-exfiltration route via satellite.[22]  In 2018, Chinese hackers ([APT10](#)) were accused of stealing intellectual property and other confidential data from satellite communications operators and others.[23]  Also in 2018, [Russia](#) was accused of jamming GPS signals during NATO military exercises, which affected nearby civilian air traffic.[24]

Only in the past couple years, the world has seen its first true "space war" in which space systems were used in an armed conflict *by both sides*.[†]  In 2022, on the day it invaded Ukraine, Russia was blamed for "bricking" or rendering inoperable [Viasat](#) modems with malware, denying access to satellite internet services to disrupt information and coordination efforts.[25]  As with many cyberattacks, its effects were indiscriminate because they're inherently hard to predict and control.  Thus, [tens of thousands](#) of other Viasat customers across Europe also experienced outages because of Russia's hack, affecting not only internet use but also remote monitoring of nearly 6,000 wind turbines, for example.[26]

Also that year and apparently still ongoing, Russia attempted to jam and hack into [Starlink's](#) satellite internet equipment, including planting malware on [mobile devices](#) to steal military communications from Starlink servers.[27,28]  But those attacks have been unsuccessful until only [recently](#), about two years later, given the novelty of Starlink's technology and its unfamiliar vulnerabilities.[29]

These Russian cyberattacks were a global wake-up call for space cybersecurity.  But this isn't to say that it wasn't a serious concern for defense planners, just not discussed much in the broader media and public, if they were aware of it at all.  A couple years earlier and presciently, in 2020, the US Department of Defense launched its "[Hack-a-Sat](#)" competition, a capture-the-flag style contest that culminated with several teams successfully hacking into a real satellite in orbit at the 2023 DEFCON hacker convention.[30]  Also in 2023, the [European Space Agency](#) (ESA) oversaw a demonstration of hacking an operational satellite.[31]

Many more incidents have happened in between the few listed above, such as the hack discovered in 2024 on South Korea's [Satellite Operations Center](#), presumably by North Korea.[32]  While they aren't as widespread (yet) as other terrestrial cyberattacks—the daily barrage suffered by industry, state organizations, and individuals that is all too normalized— it's becoming an increasingly urgent and dangerous problem for several reasons, including the following.

---

[†] The US-led [Operation Desert Storm](#) (1991) in the Gulf War has also been called the "first space war" because it involved using GPS and satellite communications against Iraq; but Iraq did not target those space systems nor use any in its own operations.  That seems like a "space war" since *both* sides didn't have space capabilities.  Similarly, a war between a modern military and an indigenous culture without guns wouldn't properly be a "shooting war" if only one side were shooting.

### 1. Space race 2.0

Outer space is rapidly becoming more congested and therefore more contested, given ongoing geopolitical tensions on Earth. As tracked by the [chart](#) below, the trending launch numbers are nothing short of astonishing.[33]

According to data from the United Nations Office of Outer Space Affairs, for nearly the 50 years between 1965 to 2012, the total number of registered space objects (primarily satellites) launched worldwide has held steady, averaging around 130 per year. But that suddenly changed with an average of more than 220 space objects from 2013 to 2016, which then *doubled* to nearly 500 objects from 2017 to 2019.

That figure *tripled* over the two years to more than 1,500 objects from 2020 to 2021. And the world is on pace to double that before long, launching an average of nearly 2,600 objects in the last two years, 2022 and 2023. To be sure, most of this activity is from the United States, but competitor states are paying attention and catching up.

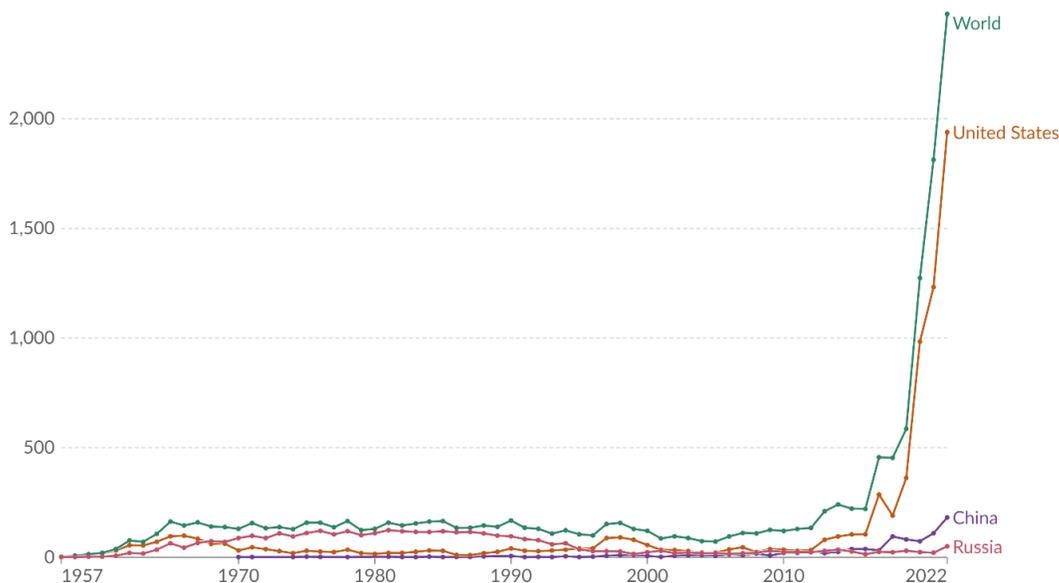

**Annual number of objects launched into space**
This includes satellites, probes, landers, crewed spacecrafts, and space station flight elements launched into Earth orbit or beyond.

**Data source:** United Nations Office for Outer Space Affairs (2024)
**Note:** Where they differ, launch attributions are based on the commissioning country, not the country conducting the operations.

*Chart 1: annual number of objects launched into space. Credit: OurWorldInData.org*

This **exponential growth** in space launches is partly driven by new, more capable technologies, notably artificial intelligence (AI) and robotics.  For instance, SpaceX has developed a reusable first-stage rocket booster that can land back on Earth with precision, which is reportedly already saving 60% of the typical launch price.[34]  With the decreasing size of computers, small satellites are also making outer space more affordable and therefore accessible.  A popular example is the CubeSat, which starts at 10 cm x 10 cm x 10 cm, less than twice the size of a Rubik's cube—affordable enough for even academic labs to launch.[35]

These smaller, more capable satellites can be coordinated with AI to form constellations or "flocks" in orbit, such as to provide satellite phone services (e.g., Iridium), satellite internet (e.g., OneWeb), and satellite imagery (e.g., Planet Labs).[36]  No longer are satellites mostly just bent-pipes for relaying radio signals through orbit or beaming sensor data down to Earth, but now with on-board computing and increasing autonomy, e.g., for flight control, they present a wider surface of capabilities for hackers to compromise or exploit.  That is to say, threat actors can do more, once inside a space system.

With greater access and capabilities in space, the global competition for space resources and research sites is heating up.  For instance, the South Pole of the Moon is a highly prized region for its scientific research value, in addition to the possible abundance of ice.[37]  The significance of lunar ice is that it can be converted into water for astronaut consumption and other purposes, such as a coolant or even fuel (after electrolysis to separate hydrogen and oxygen atoms); but given its weight, water is expensive to launch from Earth, especially in quantities for extended and long-range missions.  The Moon could become a waypoint to replenish water, oxygen, and other supplies before heading further into space.

In 2023, Russia's lander crashed en route to the South Pole of the Moon, ending in mission-failure.[38]  Only a few days later, India became the first nation to successfully soft-land a spacecraft in the region.[39]  Then in early 2024, the US reached the region with the first successful soft-landing of a *commercial* spacecraft on the Moon.[40]  Not to be outdone, China is now planning to land in the region in 2026 as the foundation for its planned lunar base.[41]

Not just state actors, this increased accessibility means that companies can also afford to play in outer space.  However, what little space law exists internationally doesn't provide much help to govern commercial activities in space, which is left to each state to regulate as they see fit.  As discussed below in this section, on legal regimes, the core texts in space law are 50+ years old, drafted at a time when it was believed that only states could afford to launch into space; as a result, space law focuses primarily on states, not industry or other private actors.

As a consequence, space commerce has seen plans that perhaps lack the dignity that the Moon and the cosmos deserve. These include sending human remains to the Moon over cultural and religious objections, as well as a marketing stunt to send powdered sports-drink that future astronauts could rehydrate with lunar water; both were part of a single Astrobotic launch that failed in 2024.[42] Future schemes include planting a Christian cross made out of regolith or lunar soil on the Moon, which can be expected to provoke even greater cultural and religious outrage—also planting the seeds of discontent and conflict.[43] These new controversies and players may surface new attacker motivations and personas that haven't historically been considered in space security.[44,45]

> *Space commerce has seen plans that perhaps lack the dignity that the Moon and the cosmos deserve.*

And where space science and security were behind the push into space in the first 50 years of space exploration and driven by states, notably the US and former USSR, **economics** is now also a driver for competition and therefore possible conflicts. For instance, states and industry are very much interested in extracting and exploiting space resources, from ice to precious metals. Consider Asteroid 16 Psyche, which orbits the Sun between Mars and Jupiter: it's larger than the island of Cyprus and full of iron and nickel, estimated to be worth around $10 *quintillion* (18 zeros)—orders of magnitude larger than the entire world's economy, which is worth about $105 trillion (12 zeros).[46] This asteroid is one of many potential treasure chests in space that may spark a new gold rush of sorts.[47]

## 2. Remoteness of space

Outer space is very far away—and therefore is very expensive to access—so much so that fewer than 700 people have ever been in space.[48] Compared to the current world population of over 8 billion people, that's essentially a rounding error. There are more than four times the number of billionaires in the world.[49] Thus, the primary way we interact with space is digitally, e.g., through our mobile phones, satellite dishes, and websites with space-data feeds. It would stand to reason, then, that if someone wanted to attack a space asset, they must also do it digitally, that is, with a cyberattack; remote attacks don't depend on physical proximity or access.

Certainly, a handful of nations—the United States, China, Russia, and India—have already demonstrated anti-satellite (ASAT) capabilities, shooting missiles from Earth to destroy targets in orbit.[50] But calls to ban anti-satellite testing (for direct-ascent "kills", e.g., ground-based projectiles, but not for other means, such as via cyberattack) by both the United Nations and the US have enjoyed much momentum in the last couple years; the 2022 UN

General Assembly resolution on the same passed in a vote of 154-to-8, with notably China and Russia among the dissenters.[51]  Therefore, any nation that conducts direct-ascent ASAT would do so at the great risk of international condemnation, so that seems to be an unlikely scenario in the foreseeable future, or so the world would hope.  (See also the next point for a related reason.)

The vast physical distance from Earth to space isn't just a challenge for attackers but also for defenders.  Consider that, right now, the oldest *functioning* satellite in orbit is Lincoln Calibration Sphere 1, which was launched in 1965.[52]  (Vanguard 1 is the oldest satellite in space, launched in 1958, but it stopped functioning in 1964 when communications with the satellite was lost.[53])  But modern cybersecurity as a concern didn't even exist until the 1970s at the earliest with Creeper, the first computer worm that was unleashed on ARPANET, the precursor of today's internet; and even then, cybersecurity didn't start becoming a mainstream concern until the late 1980s, growing to epidemic proportions in the 1990s alongside the rise of the internet.[54,55]

This means that, right now, there are working satellites in space with *no defenses* built in, since they pre-date serious concerns about modern cybersecurity.  Again, the first publicly known incidents of satellite-hacking, beyond signals jamming, didn't occur until at least the late 1990s (ROSAT and Skynet) or early 2000s (Landsat-7 and Terra AM-1), depending on which accounts you believe.  And, as crazy as it sounds, satellites are still being launched today with *no* cybersecurity, such as CubeSats that are popular with university labs and others for their inexpensive cost to build and launch; they typically have neither the onboard room to squeeze in cybersecurity components nor the budget for it anyway.[56]

Even for satellites and other spacecraft that have cyber defenses built in, we all know by now that cybersecurity is an unending game of Whack-a-Mole.  It's not enough to have the latest technology and operating system because they all have inevitable bugs somewhere within their millions of lines of code, written by entire teams of (fallible) human programmers with no single person able to understand the full picture.[57]  As a result, we need to continually upgrade our hardware, not just for faster and more powerful performance but also to fix previous and new vulnerabilities.  Software also requires continual patches, until the next version of the operating system or app is released.

But these upgrades, patches, and maintenance are enormously hard to do with objects in space.  The physical distance from here to any orbit, not to mention the effort and cost to get up there, means we can't just swap out a server anytime we like as we might on Earth.  Space cybersecurity, then, needs to make do with what hardware is on board, which sets limits to how far firmware and software can be upgraded remotely.  Major system updates are typically avoided to remove the risk of a failed update that could brick a spacecraft, and there are generally permanent software components that cannot be patched remotely. All this cyber insecurity worsens over time.

Further, in orbit and other parts of the cosmos, space objects have nowhere to hide; they're plain to see by anyone with the right telescope or other means. And because orbital mechanics follows the laws of physics and is predictable, the movement of space objects are also fairly predictable, apart from deliberate maneuvers and uncontrolled collisions or "conjunctions." This can make them particularly vulnerable as targets that can be tracked and in one's line of sight.

By the way, space weather becomes a more serious factor to consider when going off-world, for both humans and our technology.[58] The harsh space environment includes extreme temperatures, radiation (e.g., from solar flares), and impact from micrometeoroids that exceed the normal operating limits for hardware on Earth. For instance, space radiation can cause bit flips—turning 1s into 0s, and vice versa—that corrupt the storage of cryptographic keys onboard, which could cause a valid credential to cease working and deny access. So, special care is needed to guard against these natural hazards, which can degrade cybersecurity defenses, and some cyberattacks could possibly expose a spacecraft to these hazards.

Major space systems, such as the International Space Station (ISS) and GPS satellites, require a decade or more to develop and so are challenged in staying on top of best practices in cybersecurity from conceptualization to launch. Thus, without the latest cyber defenses—with some systems 10 years behind in cybersecurity upon launch—or much ability to stay on top of cybersecurity given hardware and software limitations, space assets represent particularly tempting and high-profile targets for attack.[59] And the primary mode of attack will likely be remote, that is, through cyberattacks or electronic warfare (e.g., signals jamming and spoofing) given the physical inaccessibility of outer space.

### 3. Space debris and sustainability

The environmental benefit of a cyberattack versus a kinetic attack is enormous in outer space. Cyberattacks typically don't blow up targets, though sometimes they can. Armed conflict is both a moral and environmental disaster, so preventing environmental harm would be a positive development, whether on Earth or in orbit. Space debris, as explained here, is an indirect factor in the rise of space cyberattacks.

Space debris is the junk we leave in space, such as dead satellites, spent boosters, fragmented pieces of rockets, broken solar panels, lost tools, and so on, some of which were created from earlier collisions or conjunctions between those space objects.[60] The problem is that they're not just floating in space peacefully; they're zipping around the globe at incredible speeds of 20,000 miles (or 32,000 kilometers) per hour or more— or, as a more conceivable number, 6 miles (or 10 km) per second or even faster.[61]

This means the space domain is already in a precarious state with respect to orbital junk. Thus, it's critical to track as many of those things as possible in order to coordinate space traffic, such as to know that we're not launching a rocket into a debris field, or to know when and where to maneuver a satellite away from a collision course with debris or even another satellite. At those velocities, any collision can be catastrophic, even with something as tiny as a fleck of paint.[62]

Space agencies are able to track about 35,000 space objects, only about 9,000 of which are working satellites and the rest is literal junk.[63] But there are also over 1 million small bits of debris that are too difficult to track, ranging from 1 to 10 cm in size.[64] Worse, there are over 130 million bits of space debris under 1 cm which our current technology cannot track.[65]

What does this have to do with cybersecurity? The threat of space debris is a powerful deterrent against a kinetic conflict in space, which would create more debris. This isn't just an environmental concern but a self-interested one: space junk doesn't care about the nationality of what it's crashing into, and it's just as likely to destroy one's own space assets as it would an adversary's. But space cyberattacks would be much less likely to contribute to this debris problem, which is in everyone's interest insofar as they benefit from space services.

> *Space cyberattacks can avoid the space debris problem and therefore may become the primary mode of conflict in orbit.*

Related to the previous point, calls for an international ASAT ban are aimed at preventing an arms race in outer space, but they're also to protect space sustainability, given the shared problem of orbital debris. Various estimates have blamed direct-ascent ASAT tests for creating 25% more trackable space debris and doubling the risk of a catastrophic collision with a spacecraft.[66,67] This by itself already suggests that a kinetic, "shoot 'em up" battle in orbit could very well be a **Pyrrhic victory** at best—that is, suffering losses so great (to any side) that even winning a space battle feels like losing—and is therefore irrational and unacceptable.

The risk from space debris isn't only about conjunctions with individual spacecrafts, but there's a planetary-scale risk. This is the "Kessler syndrome" that so many experts are worried about—a chain-reaction of uncontrollable, unpredictable collisions in the orbiting fields of space debris that generate even more debris with each collision, which means that no one will be able to safely launch off-planet without passing through this dangerous minefield that cocoons the globe.[68]  In a worst-case scenario, this effectively puts an end to all space programs on Earth, as well as the space services that enable the modern world. Researchers are also worried about the disruption caused by orbital debris, which is mostly metal, to the Earth's protective magnetic field.[69]

So, compared to kinetic conflicts, space cyberattacks can avoid the space debris problem and therefore may become the **primary mode of conflict** in orbit.

## 4.  Complexity of systems

Space systems can be extraordinarily complex, which are really "systems of systems", and this could translate into more cyber vulnerabilities, including a wider attack-surface or points of entry for a hacker.  First, the space ecosystem itself is complex and unique, considering the various segments that need to be secured: launch, ground, space, user, and communication link segments, as described in the previous section.

Looking only at the space segment as an example, think about the diverse forms of technology that are being launched into space: rockets have propulsion systems, navigation systems, as well as different payloads depending on the mission, such as satellites, space probes, crewed spacecraft that require life-support systems for humans and animals, cargo such as autonomous rovers and space telescopes, and more—all of which depend on computing systems, and all computing systems can be hacked.

The attack-surface widens with a long and distributed supply chain which represents more opportunities for errors and vulnerabilities to arise, particularly where different systems need to be integrated.  This is basically the leaky cybersecurity problem that can face an Internet of Things (IoT) environment, especially when different vendors are involved.[70] Popular in CubeSats and other satellites, commercial off-the-shelf (COTS) components contribute to this problem, because having an easy plug-and-play system comes at the expense of strong or at least consistent cybersecurity across the entire system.  For instance, the same exploit could be used over and over against different targets that use the COTS component.

While recent progress has been made toward a clear standard for space cybersecurity, such as from NASA, NIST, and IEEE, it is still a moving target since both offense and defense will continue to co-evolve, and that progress still does little for space objects previously launched with less-than-ideal cybersecurity.[71,72,73]  Moreover, standardizing space security can be difficult because satellite mission security depends on the security properties of a great range of semi-independent systems and software—that is, a system of systems.  On the other end of the spectrum, small CubeSats again typically have no cybersecurity at all, given a lack of onboard room and funds, especially for academic projects.

Nation-states are no longer the sole or dominant player in outer space, given the falling cost of launches and strong commercial interests in space.  That adds to the challenge of setting a cybersecurity standard across the supply chain, particularly as space technologies are considered to be dual use, i.e., to have both defense and non-defense applications.  While defense and intelligence missions can be expected to use higher cybersecurity standards than industry and academic missions, even those classified missions can involve commercial vendors that may have different levels of attention to cybersecurity, assuming those companies even have cybersecurity experts on staff.[74]  For instance, different data exchange interfaces between defense and civilian sectors can create gaps in information security.[75]

And we've just been talking about only the space segment in the above.  Other segments have associated technologies that need to be secured and which also involve computing systems, such as launch facilities, ground stations for communications, mission control centers, end-user terminals, and so on.

New technologies are driving the rush into space today, such as reusable launch vehicles and more capable AI for greater control and maneuverability.[76,77]  Many are prototypes that have never been used before.  This novelty brings with it the virtue of being an unfamiliar and therefore more difficult system for hackers to exploit, which helps to explain why it took two years for Russia to disrupt Starlink's satellite internet services to Ukraine.[78]  But novelty also cuts the other way; by definition, new technologies are under-studied with respect to their cyber readiness, and classified technologies don't benefit from the scrutiny of a broader range of experts, such as academic researchers.

Given a general lack of public information about technical details—whether it's from novelty, security classification, or the technology developer's unwillingness to reveal details, such as satellite firmware—certain forms of cyberattacks may require enormous resources to plan and execute, perhaps state-sponsorship as Stuxnet did.  But this "security by obscurity" can take us only so far, and attackers can target weaker links in the chain, such as with ordinary social-engineering tactics to induce space systems operators to reveal their log-in credentials (just one may be enough, out of a targeted organization that can employ thousands), or signals jamming which doesn't require any knowledge of the inner workings of a GPS system, other than its radio frequency.  Anyway, "security by obscurity" might not be viable for long, given commercial pressures to use COTS and standardized components.

While modern technology is increasingly complex and enables greater capabilities, complexity can also create brittleness, more points of failure, and cyber vulnerabilities.  And space systems are some of the most complex technologies we've created. For highly skilled hackers, breaking into a space system can be a prized trophy given the intense security surrounding launches, space missions, and their technical systems; this was the motivation of the "Hack-a-Sat" competition that resulted in several winning teams in 2023.[79]

## 5. Unclear legal regimes

Where law is unclear or lacking, ethical questions can arise in those gaps, since law often tracks morality to protect a society's values, such as a prohibition on murder and fraud. But also, ambiguity in law leaves the door open for controversial, provocative behavior. In outer space, this means potentially igniting or escalating conflicts, which can elicit a cyber response. In the following discussion, we will look at key issues in domestic law and international law, including space law, that might enable or be contributing factors to cyber conflicts in outer space.

Compared to kinetic or physical attacks, cyberattacks are not governed by a clear legal framework, especially transnational attacks and even less for space cyberattacks. Many high-profile cyberattacks—think about OPM, Sony Pictures, Equifax, Colonial Pipeline, NASA satellites, and so on—have yet to result in criminal charges or prosecutions, even where the perpetrators have been confidently identified by experts and "wanted" posters drawn up.[80] This reflects a lacuna in law or its enforcement, or even perhaps a sign of lawlessness, when serious breaches at such a scale can't be prevented or prosecuted.

Indeed, one top FBI agent suggested to "just pay the ransom" in some incidents of ransomware attacks, which isn't very helpful advice, particularly as ransomware is so rampant and extortionist.[81] As a form of self-help in the absence of any other help, the legality of "hacking back"—i.e., to hack into a cyberattacker's system to recover/delete one's stolen data or to track/disrupt their system to prevent further attacks—is also unclear as computer laws, e.g., Computer Fraud and Abuse Act (CFAA) in the US, don't typically contemplate and therefore don't address that particular scenario.[82,83]

If any given nation has a difficult time dealing with cyberattacks, then the challenge is doubled when it comes to developing *international* laws for the same, even if everyone recognizes cyber threats to be a serious problem. International discussions to develop norms for cyberspace have stalled for more than a decade over a basic lack of consensus on whether and how international law applies to cyberspace, including the law of armed conflict and international humanitarian law.[84]

And it's not even clear that cyberattacks can trigger the right to self-defense or collective defense under international law, since the "use of force" and "armed attack" prohibited by the 1945 United Nations Charter (articles 2.4 and 51 respectively) were conceived of and is still typically understood in the context of *kinetic* attacks, such as bullets and bombs, not cyberattacks.[85]  To wit, how is transmitting code—a series of 0s and 1s—a "use of force" or an armament in the first place?[86]

But if we understand "use of force" and "armed attacks" in terms of having a certain *scale, duration, and effects*, then some serious cyberattacks may arguably rise to the level that could justify a military response, such as permanently destroying a power grid that delivers energy to a civilian population, which is akin to the damage from a missile.  In contrast, having GPS signals jammed for a short period of time (which we're including as a "cyberattack" for the purposes of this report, per our terminology discussion in the previous section) seems less serious than the permanent, irreversible damage typically caused by munitions.

If we're talking about a signals attack specifically, then the International Telecommunications Union's (ITU) is relevant since it coordinates spectrum allocation for radio signals and related matters.[87]  In article 45 of its Constitution: "All [radio] stations, whatever their purpose, must be established and operated in such a manner as not to cause harmful interference to the radio services or communications of other Member States or of recognized operating agencies, or of other duly authorized operating agencies which carry on a radio service, and which operate in accordance with the provisions of the Radio Regulations."

This would seem to clearly prohibit signals jamming and spoofing, but unfortunately ITU has little enforcement mechanism and mainly relies on the good-faith cooperation of the parties involved in a dispute.[88]  Insofar as a cyberattack implies a breakdown of good faith between the parties, ITU's role is primarily in standards setting, coordination, and development, and so it provides only limited deterrence against cyberattacks, e.g., if states expect to want ITU as a neutral arbiter for future disputes and need to stay in its good graces.

*Space* cybersecurity is triply challenged insofar as they implicate international space law, which is decades old and mostly silent on cyberattacks, space debris, and other current concerns that didn't exist then.  The main instrument for space law is still the **Outer Space Treaty** (OST), ratified in 1967, well before modern cybersecurity was a concern.[89]  As a product of the Cold War, OST was meant to only be a foundational agreement, a basic framework that the US and former USSR could minimally agree to as geopolitical competitors and the only space powers of the age; so, it's not a surprise that it would leave many details open.

The three subsequent treaties in space law include the Rescue Agreement (1968), Liability Convention (1972), and Registration Convention (1976).  These treaties stipulate, respectively, obligations to render assistance to astronauts, liability for damage caused by space objects, and the registration of space objects in the interests of space traffic management (STM) and international cooperation.

Other international space law includes "soft law" instruments such as UN declarations, principles, recommendations, and guidelines.  While soft law is not legally enforceable, it is nonetheless authoritative as the expressed preference of behavior by and for states.  The Moon Agreement (1984) isn't legally binding for most states as only 17 states have ratified it to date.  Collectively, these sources of space law can serve as a basis for establishing norms that can develop into customary international law or further hard law, but as relevant to this discussion, none of them directly addresses cybersecurity.

> *The bulk of space law is generally not relevant or adequate to deal with space cybersecurity or other modern problems.*

So, besides being limited to only a few core instruments, the bulk of space law is generally not relevant or adequate to deal with space cybersecurity or other modern problems.  We will reserve a much longer discussion of that for a later publication, but for now, it's enough to point to a few key provisions in OST that appear relevant to space cybersecurity:

In OST, Article IV and others affirm that outer space is to be used for peaceful purposes only: "The Moon and other celestial bodies shall be used by all States Parties to the Treaty exclusively for **peaceful purposes**.  The establishment of military bases, installations and fortifications, the testing of any type of weapons and the conduct of military maneuvers on celestial bodies shall be forbidden" (emphasis added).[90]

Article IX establishes a principle of non-interference: "If a State Party to the Treaty has reason to believe that an activity or experiment planned by it or its nationals in outer space, including the Moon and other celestial bodies, would cause potentially **harmful interference** with activities of other States Parties in the peaceful exploration and use of outer space, including the Moon and other celestial bodies, it shall undertake appropriate international consultations before proceeding with any such activity or experiment.  A State Party to the Treaty which has reason to believe that an activity or experiment planned by another State Party in outer space, including the Moon and other celestial bodies, would cause potentially **harmful interference** with activities in the peaceful exploration and use of outer space, including the Moon and other celestial bodies, may request consultation concerning the activity or experiment" (emphasis added).[91]

Article I may also be relevant, as it lays down a principle of non-discrimination: "Outer space, including the moon and other celestial bodies, shall be free for exploration and use by all States without **discrimination** of any kind, on a basis of equality and in accordance with international law, and there shall be free access to all areas" (emphasis added).[92]

Launching a cyberattack in space, then, would seem to violate at least some of these provisions.  As examples: contrary to article IV, it would be either an aggressive, non-peaceful activity intended to exploit a computing system without authorization, or a test of a (cyber)weapon, which is also forbidden.  Contra article IX, a cyberattack would seem to qualify as prohibited harmful interference, even if not kinetic (which OST never specified it needed to be).  And contra article I, an attack that targets space assets of only a particular nationality, e.g., a jamming attack on US satellites, would seem to violate the principle of non-discrimination, as opposed to a spectrum attack that jammed signals for all satellites.  Yet, space cyberattacks still happen without any legal charges of violating OST.

But these are interpretations of OST and may be disputed.  For instance, it may be claimed that a space cyberattack was necessary to intervene with a prior aggressive, non-peaceful activity; and if that prior activity threatened to interfere with another state party's plans, then the cyberattack looks more like a case of "hacking back", the legal status of which is unclear under domestic if not also international law.[93]

Just about all space technologies and objects have the potential for dual use, i.e., for peaceful and non-peaceful uses, as is widely understood.  As such, there may be plausible deniability in launching a space system that's capable of offensive cyberoperations, which helps to conceal the weapon until it's too late.

Further, what counts as prohibited harmful interference in outer space is unclear insofar as it's untested in any legal proceeding.  For instance, interference seems to happen constantly by way of signals jamming/spoofing, and interference can also come from unintentional acts, such as forcing a satellite to maneuver away from its orbit because of a possible collision with a second satellite.[94]  A considerable part of the problem is that the source of signals interference can be difficult to identify and track, even more so when the source is mobile, such as onboard a ship.  Again, no nation has ever been found to violate OST yet, even if some disputes have arisen related to OST.[95,96]

Even if the law, either domestic or international, were clear about the illegality of cyberattacks, provided an enforcement mechanism, and specified legal remedies, a key barrier to a successful prosecution continues to be clear attribution of the attack.  For instance, though China was publicly suspected to be behind the hacking of NASA's Landsat-7 and Terra AM-1 satellites—a serious enough attack, as with any satellite hacking, that would justify legal actions to deter future breaches—there was enough [doubt] or plausible deniability to think that a legal case wouldn't be successful.[97]

But little is clear about applicable law and governance around international cyberattacks, much less those that occur in outer space.  This creates room for ambiguity and raises the likelihood of actors evading parameters for lawful activity in conducting malicious or harmful activity like cyberattacks, which risks escalation into a more serious conflict, even a kinetic one.  Again, not all cyberattacks on space systems occur in outer space since they could be aimed at the ground, users, and other segments on Earth; but the more novel scenarios offered in this report will tend to be about cyberattacks in space insofar as they're less obvious and less considered than the more familiar cyberattack scenarios on Earth.

The various legal lacunae above are also examples of why it's important to consider a full range of scenarios in law and policymaking—to anticipate and identify those gaps—as this report seeks to do with space-cybersecurity scenarios.  Other governance gaps in space law and policy, beyond the above issues focused on cybersecurity, can lead to misunderstandings, miscalculations, and ultimately conflicts.  Again, we will reserve that extended discussion for a later report.

Because laws can exist for decades or longer, space-cyberattack scenarios that seem outlandish or more speculative today might not be so in the distant future, just as OST didn't anticipate the modern problems we're facing now.  It would still be instructive to test proposed legal principles and frameworks against a full range of scenarios, including the more distant, to proactively close any gaps.

## 6.  Familiar cyber advantages

Cyberattacks can also be expected to continue in space for the same reasons why "regular" cyberattacks on Earth will continue: they work.  They can be a low-cost yet still highly effective tactic economically, politically, and environmentally.

To be sure, conducting digital reconnaissance or forensics to study a target's computing system for cyber vulnerabilities, such as zero-day exploits (undisclosed bugs that haven't been used before in a cyberattack), and then developing a cyberweapon to penetrate that particular system can require massive resources and entire teams of technical experts, which are expensive.  For instance, on the high end as the world's first cyberweapon, the 2010 Stuxnet worm was a very specific malware tailormade for the very specific mission of destroying the centrifuges in Iran's nuclear program, and it was estimated to [cost] "hundreds of millions of dollars" and [two or three years] to develop.[98,99]

But the barriers to access cyber capabilities continue to lower, and many other cyberattacks can be much cheaper to develop and launch than the cost of physically sending in a strike team, materiel, transportation, and so on in preparing for and waging a kinetic attack.  As computing becomes cheaper and more powerful every year, the cost of launching a cyberattack is now ridiculously affordable, as low as $30 per month for a phishing campaign, which is less than the hourly wage of the average US Navy SEAL, never mind the cost of an entire strike team over the duration of a mission.[100,101]  In contrast, launching a single missile can cost more than $2 million, though exploding or "suicide" drones as seen in Ukraine only cost a few hundred dollars.[102,103]

Though they can be cheaper, cyberoperations can be incredibly effective against modern adversaries that depend on digital systems, particularly for command, communications, control, and intelligence (C3I).  Where a drone attack could take out an entire building, the digital infrastructure for C3I isn't necessarily housed in any single, physical location but could be distributed and even virtual, e.g., cloud computing.  Therefore, even bombing an adversary's communications center into oblivion might not even be as effective as a cyberattack that transcends physical space and limits, just as its digital target does.

Cyberattacks can also offer a range of new and useful options, other than to kinetically destroy a target.  A cyberattack can be functionally equivalent to a kinetic attack if it renders the target completely useless, but it could also be designed to have reversible or temporary effects.  Or it might not be an attack at all but an intelligence-gathering operation, e.g., to eavesdrop on enemy communications.  Or it could be used for deception, e.g., to cause a compromised system to appear normal but actually provide false data to decisionmakers.

Politically, the costs of a cyberattack are much less dramatic, compared to the risk of putting "boots on the ground", especially in a foreign land which could provoke a diplomatic crisis or, worse, start a war.  Since the correlative risk of "boots on the ground" is "bodies coming home in a bag", a cyberattack is unlikely to have such a high human-cost (to our side) such that it can erode public support for the military action and threaten the political careers involved.  For space conflicts, the kinetic analogue would be to kill a satellite with a missile from Earth, which is not only dramatic and expensive but also counterproductive to hard-won progress to ban ASAT tests and incidents.

Cyberattacks can also be less provocative and therefore less escalatory because of their stealth nature.  First, it's often unclear who the attacker is, given techniques to hide their location, as well as plausible deniability since most people wouldn't understand digital evidence in the first place.  For instance, a state-actor can simply disavow its cyberattack, which may be enough to create doubt in the global public's mind.  It can also be unclear who the intended target is, as some cyberattacks can spread indiscriminately and affect thousands of other machines or more, such as malware from email or an infected website.

Second, even if the intended target is known, it's often unclear what damage a cyberattack has caused and therefore what a proportionate response would be. The effect of a cyberattack isn't always obvious, such as taking down an entire computing system, but it could be more surreptitious, such as stealing data, planting bad data, or installing spyware to eavesdrop on communications. Since some cyberattacks, such as zero-day exploits, have never been attempted in the field before, as opposed to a controlled lab setting or digital sandbox, it's not at all assured that the cyberattack will even work as intended. It may have partial or unintended effects which can be worse than expected and therefore escalatory in that case, e.g., if an entire power grid to civilian populations were taken offline that resulted in deaths in the winter cold or summer heat.

Third, it's often unclear that a cyberattack has occurred in the first place, as that can look like a normal, non-cyber failure of a system. In its 2022 report, IBM estimated that, on average, it takes an organization 207 days (about 7 months) to identify that a breach has occurred, and then another 70 days to contain it.[104] If it takes months or years to discover that a cyberattack had even happened, that lag-time cuts against the urgency and impulse for swift action, even if the perpetrator is definitively identified.

The risk is worse in space since access to monitoring telemetry may be sporadic and limited to periods of line-of-sight. Unless remote data forensics is built into the spacecraft by design, which is rare, conclusive forensics may require physical access, such as to disk artifacts, and so is not a practical option given that outer space is so difficult to access.

On top of all this, many cyber breaches go unreported for a range of reasons, including not wanting to give an adversary or hacker a damage assessment of how effective a particular cyberattack was. If the attack was effective, publicly disclosing that may encourage more such attacks; and if the attack was ineffective, it would waste the attacker's time to continue attempting the tactic, as well as provide more digital evidence about the perpetrator. Public reporting can also scare away customers and investors, among other self-interested reasons.

As a result, we can reasonably expect that outer space cyberattacks will be an attractive and effective tool for adversaries and other bad actors for at least the same kinds of reasons that cyberattacks are so widespread on Earth.

## 7. Higher stakes

The final reason we'll discuss here for increasing attention to *space* cyberattacks is that the stakes are unusually high, as compared to other cyberattacks, and more than most people may realize; therefore, the threat needs more attention.

First, many critical services are enabled by space systems. As mentioned in the introductory section, GPS and other satellites provide positioning, navigation, and timing services that we use daily. Weather and other Earth-observation satellites provide data and imagery for monitoring, managing, and forecasting many things we care about, with agricultural, ecological, social, economic, and political (national security) benefits. So, a loss of those services would be highly disruptive, for instance, to air and marine navigation as well as financial transactions and other communications. Moreover, space systems are seen as "a direct enabler of military operations" and therefore are highly strategic targets themselves.[105]

*Even if important services aren't affected, space systems can be high-profile, symbolic targets.*

Even if important services aren't affected, space systems can be high-profile, symbolic targets. These may be more important to, say, terrorist and activist groups who are looking to shock with a spectacle, as opposed to state-actors who are looking to cause confusion and disrupt information as a prelude to a kinetic attack or some other action. And given that cyberattacks can be much less costly and risky than a kinetic attack, non-state actors can better afford to launch cyberattacks to cause any number of effects, from the trivial to serious. If a space system is behind in its cybersecurity measures, that makes for an even more attractive target.

Constellations of coordinated small satellites are increasingly popular for their massive coverage areas, such as to beam down satellite internet, phone, and television signals, but they could have unique cyber vulnerabilities. For instance, each satellite in a swarm or constellation is typically identical or near-identical to the others; this means that a cyber exploit that works on one satellite could work on them all. Where it would be most impractical to take down a constellation one satellite at a time, e.g., each with its own very expensive ASAT missile, it would be much more plausible and cost-effective to develop a single zero-day exploit for the job. Further, these constellations raise new problems, such as creating more light pollution that interferes with astronomy, in addition to possibly adding more orbital debris given the sheer number of satellites involved.[106]

Space debris continues to be one of the most urgent threats in outer space, at least in the orbits we care about. While cyberattacks have the virtue of typically *not* generating more space debris, that may not always be the case. Stealing data from a satellite, for instance, wouldn't disrupt the satellite's operations, but a more serious cyberattack could disable a space object completely, turning it functionally into a single, large bit of space debris that could become many more bits later.

And depending on what system is targeted within a space object, a cyberattack can result in a kinetic effect—such as overheating and overloading batteries that lead to an explosion—which can fragment the space object into many bits of space trash, potentially thousands, that also need to be tracked and avoided.  A disabled space object and its components can also come raining back down on Earth if it deorbits, potentially causing harm to property and life, either unintentionally or by design.

While cyberattacks can be less provocative and therefore less escalatory because of their stealth nature and lack of mass casualties, as discussed above, the special circumstances of the space environment could also elicit the opposite response.  Every nation already has trouble in preventing cyberattacks on Earth, and there's reason to believe cyberattacks in space will increase, also discussed above.  Meanwhile, nations are highly motivated to deter cyberattacks on their systems, especially those critical to national security and intelligence.  Without spy satellites and other capabilities, modern militaries would be at a great disadvantage without a vantage point in orbit and other space-enabled services, such as GPS and communications.

This confluence of factors have apparently led Russia to declare that any hacking of satellites could be received as an act of war or *casus belli*.[107]  Whether the threat was hyperbolic or real, it still has the intended effect of deterrence, since even the possibility of war would be a frightening specter that can move global financial markets and cause other chaos.  It's also disconcerting because war is supposed to be a last resort since it's so terrible, yet Russia is signaling its willingness to slip down this path over mere satellite hacking—in which no one dies or is even harmed, at least directly—which has yet to provoke even a criminal prosecution in its short history between nation-states, much less such dramatic saber-rattling until now.[108]  This threat also underscores the essential role that space assets play in modern militaries, as well as the desperation that Russia and all other nations have in securing space systems from cyberattacks.

And war with Russia could be existential for everyone.  Russian president Putin and others have already signaled a **scorched-Earth strategy** in case the country comes under serious attack, issuing such remarks such, "Why do we need a world without Russia in it?"[109,110]  Again, whether or not this threat is a bluff, it can still have the desired deterrence effect.[111]  Besides imposing an outrageous, outsized cost to its adversaries, the actual or perceived unreasonableness disrupts calculations in game theory which attempts to plot a winning path or moves based on expected reactions or moves from the adversary or competitor; an irrational actor tends to defy expectations.

On brand with its scorched-Earth strategy and as of this writing, Russia could be desperate enough to back out of the Outer Space Treaty, which it had negotiated and ratified in 1967, if it really intends to send a nuclear weapon into orbit, even as an anti-satellite weapon only.[112] Indeed, Russia is starting to explicitly signal that it may be ok with the use of nuclear weapons in outer space.[113,114]

If Russia is not bluffing, such a move would break one of OST's fundamental principles; in Article IV: "States Parties to the Treaty undertake not to place in orbit around the Earth any objects carrying nuclear weapons or any other kinds of weapons of mass destruction, install such weapons on celestial bodies, or station such weapons in outer space in any other manner."[115,116]

In case nations still exist after a major kinetic war on Earth, the higher stake here with space cyberattacks is related to the problem of space debris. A kinetic war on Earth, especially between space powers, can spread into a kinetic war in space, which is not prohibited by any legally binding rules. If a single direct-ascent ASAT test can generate 25% more space debris than what exists—and neither ASAT tests nor debris creation are legally prohibited—then a full-blown battle in space would be devastating to everyone's space ambitions by ruining orbits for hundreds or thousands of years.[117,118,119]

While the risk of the Kessler syndrome—a planetary prison made from orbital debris—can and should be a deterrent for aggression in outer space, it could also be the *goal* of some cyberattackers, whether irresponsible nation-states, terrorists, hacktivists, or other actors, as posited by some of the scenarios in this report. For instance, a non-spacefaring state with no space assets and no reliance on space services, at least in its own (mis)perception, might conclude it has nothing to lose in poisoning Earth's orbits, though how such a state might secure ASAT missiles or the cyber expertise needed would be an open question.

.....................

As a concluding note to this section: even if cyberattacks are the dominant form of conflict in outer space, and costs of a kinetic conflict could be unacceptably high for rationale actors, this does not mean we don't need to urgently develop more effective space governance and incentives to prevent such kinetic conflicts and arms races.[120]

The ongoing militarization of space, with few guardrails, is still concerning to many, and miscalculations could accidentally lead to physical battles in space.[121] For instance, projects to give satellites the defensive ability to autonomously track, disable, or destroy other satellites need to seriously consider their impact on the environmental sustainability of orbits as well as on escalation dynamics, especially in the event of a wrongful defensive decision by either an AI or human and in the event of a conflict with adversarial satellites in the future that also have autonomous defenses.[122]

Wars typically involve some *unreasonable* actor that would threaten millions of human lives, along with economic and environmental damage, for questionable gains, so we cannot rely on rationality or luck to protect the fragile space environment from human wars.  Even limited-scale kinetic conflicts, even a single incident, could quickly escalate into a wider conflict in space and/or create new debris that has cascading effects, pushing us closer to the Kessler syndrome.  At the least, a limited-scale kinetic conflict in space would set a dangerous precedent and perhaps lower barriers for the next kinetic battles, once the proverbial floodgates are open.

Beyond the governable state actors, cyberattacks can come from non-state actors who might not care at all about space sustainability, e.g., chaos agents and other extremists, who would be happy to bring about the Kessler syndrome.  For instance, their cyberattack might target a critical component of a spacecraft, the failure of which could result in an explosion and thousands or more pieces of dangerous space junk.

And to the extent that cyberwarfare can be governed, the governance of *space* cyberwarfare should of course be addressed, too.  The challenge, however, is that terrestrial cyberwarfare and cyberattacks have been stubbornly resistant to international agreements, though the special circumstances of outer space may provide hope for better progress.[123]

With the above context, in the next sections we will introduce the ICARUS matrix to generate novel scenarios in space cybersecurity, along with a robust set of novel scenarios and critical-thinking questions to develop and interrogate new scenarios.

## 03

# Taxonomy: how to generate novel scenarios

Humans are incredibly inventive and smart when we want to be, for better and worse. In cybersecurity, this means an ability to continually surprise the world with new ways to penetrate and exploit a digital system. As a result, cybersecurity remains a persistent concern in modern life, particularly in outer space, as the previous section explains.

To guard against a particular cyber threat or attack-vector, the threat first needs to be known in order to assess if existing defenses are enough or need to be bolstered. But in this ongoing evolution between hunter and prey, cyber defenders can never relax. Both complacency and a failure of imagination can open the door for cyberattacks that cause massive economic losses and worse.

This section provides help for the imagination piece of the puzzle. Below is a new taxonomy of sorts, a scenario-prompt generator we have developed to help prime the imagination pump. This is called the **ICARUS matrix**, an acronym for "**I**magining **C**yberattacks to **A**nticipate **R**isks **U**nique to **S**pace." While there are many excellent taxonomies about cybersecurity—such as notably from NIST, MITRE, and Aerospace Corporation—they weren't sufficient for our purposes here for several reasons as follows.[124,125,126]

Some taxonomies are so technical that only cybersecurity practitioners can understand them, which is fine for some purposes; we wanted a taxonomy that's also accessible to policy planners and other experts since cybersecurity is also a policy and social problem. Some taxonomies were comprehensive and offered a great level of detail, while others were too simplified for our purposes; we wanted to strike a balance between the two. Also, general taxonomies don't capture the specific segments in the space ecosystem, e.g., space and link segments, which seem important to call out in anticipating novel cyberattacks.

But more than just a list of possible attack-vectors and types of cyber exploits, we wanted a framework that could push our imagination to be more precise about cyberattack scenarios. There can be many different motivations, methods, and so on involved, and it may be important to consider not just where the vulnerability is but from which direction attackers might be approaching it. This means we wanted a framework that could account for not only *how* a digital system can be breached, but also the *who, what, where, when,* and *why* of a scenario. All that can help point to different countermeasures and more response-options, as well as add richer details to a scenario.

Again, a key motivation for this report is that, at least in non-classified discussions, only a few scenarios are typically trotted out in space cybersecurity, namely something vague about satellite hacking as well as jamming or spoofing of signals, such as GPS. If those were the only scenarios that are being guarded against, then the entire space sector is headed for trouble. Thus, progress in space cybersecurity can greatly benefit from an infusion of many more scenarios, particularly surprising ones. As unclassified material, they would enable countless researchers without security classifications to engage with the scenarios.

With that introduction, below is the ICARUS matrix, usable by both technical and non-technical people, to seed the basic structure of a novel space-cyberattack scenario that can be fleshed out with more details; see table 1 below, or Appendix A for a larger image. While it's not meant to be a formal taxonomy per se, since many of the variables can overlap and are not mutually exclusive as with a typical taxonomy, it can be useful as a general taxonomy as it captures the major variables, and other variables can be added to it as desired.

| | A: Threat actors | B: Motivations | C: Cyberattack methods | D: Victims / stakeholders | E: Space capabilities affected |
|---|---|---|---|---|---|
| 1 | Major space-faring states | Nationalism | Insider attack | Major space-faring states | GPS / GNSS |
| 2 | Other space-faring states | Dominance / influence | Social engineering | Other space-faring states | Earth observation / remote sensing |
| 3 | Non-space-faring states | Financial / economic | Ransomware | Non-space-faring states | Military intelligence and capabilities |
| 4 | Insider threats | Fraud | Honeypot | State-owned entities | Spacecraft, robotic or crewed |
| 5 | Political terrorists | Employment | Sensor attack | Military and other contractors | Life-sustaining services |
| 6 | Mercenaries | Blackmail / coercion | Signals jamming | Scientific organizations | Other essential services |
| 7 | Eco-terrorists | Terror | Signals spoofing or hijacking | Corporations | Other safety of personnel / others |
| 8 | Corporations | Warfare | Eavesdrop / man-in-the-middle | Wealthy individuals | Loss of sovereignty / control |
| 9 | Mobile service providers | Disinformation | Network security | General population / society | Earthbound services |
| 10 | Launch service providers | Espionage | Supply chain, hardware | Indirect / secondary stakeholders | Emergency services |
| 11 | Social engineering groups | Sabotage | Supply chain, software | Marginalized populations | Financial transactions |
| 12 | Organized crime | Extremist ideology | AI / ML / computer vision | Social movements | Mining or manufacturing |
| 13 | Chaos agents | Cult of personality | Attack coverup | Cultural / religious groups | Scientific capability / research |
| 14 | Religious / apocalyptic | Paranoia / anti-technology | Software hacking | Unions / labor reps | Asteroid detection systems |
| 15 | Other ideological groups | Boredom / trolling | Systems security | Customers / users via their data | Space weather monitoring |
| 16 | Proxies / agents, esp. unwilling | See world burn / chaos | Multi-phase attack / APT | Individual targets | Space traffic management |
| 17 | Noncombatants, esp. unwilling | Social / distributive justice | Cloud hacking | Critical specialists | Space tourism |
| 18 | Amateur hackers / enthusiasts | Intellectual / tech demo | Account compromise | Critical infrastructure | Launch capabilities |
| 19 | AI / machine learning | Revenge / retaliation | Quantum computing / comms | Internet / media / entertainment | Communications |
| 20 | Unknown / anonymous | First contact, for and against | Death by 1,000 cuts / long game | AI / machine learning | News / social media |

*Table 1: ICARUS matrix. See Appendix A for a larger image.*

## Understanding the ICARUS matrix

First, ICARUS as a scenario-prompt generator does *not* offer a comprehensive list of variables but only a robust starting list to help spur "imagineering" of scenarios. There are undoubtedly more variables that can be considered, but we've limited ourselves to the major ones here as an initial offering. Moreover, new cyberattack methods, new space capabilities, new threat actors, and so on will emerge over time, especially as space and cyber technologies continue to evolve.

To that end, we identify five major categories of interest (columns A-E) for *any* scenario in space cybersecurity. The basic idea is to select one of the 20 variables (rows 1-20) from two or more of the columns to create a prompt or basic structure for a novel scenario, and the five columns represent the following major elements:

    A. Threat actors or agents (**who** is perpetrating the cyberattack?)
    B. Motivations (**why** are they launching a cyberattack?)
    C. Cyberattack methods (**how** would the attacker penetrate a system?)
    D. Victims or stakeholders (another **who** question)
    E. Space capabilities affected (**what** is the damage or effect intended by the attacker?)

It should be noted that **where** the cyberattack takes place and **when** the cyberattack might occur aren't included in the ICARUS matrix, as those aren't so much variables but *dependencies* of a scenario. For instance, a cyberattack on Voyager 1 must occur beyond our solar system because that's where the satellite is. Or a hack of the water system of a Martian colony likely wouldn't occur in the near future or even medium-term (inside 20 years or so) because Martian colonies aren't projected to exist until around 2050, given space capabilities today.[127]

In any case, the next section of this report will organize a sample set of more than 40 scenarios along the axes of time (*when* on the time-horizon could this cyberattack occur?) and place (*where* in space could this cyberattack occur, from Earth to interstellar space?). But the ICARUS matrix can create many more scenarios; there are **4,084,000 unique combinations** of variables—that is, scenario prompts—that can be drawn from the 2-5 categories or columns.[†] And there can be many more than the 20 variables we selected for each category, which again is only a starting list.

---

[†] Note that some combinations may be nonsensical, though the exercise of identifying those and explaining why can also be productive. For instance, terrorists wouldn't likely cover up their attack, or else it couldn't create the spectacle and instill the fear desired. Or an attack on space tourism

Why draw from two or more categories?  Because just as two points in geometry are needed to describe a line, we need to nail down variables from two or more categories to provide a minimally viable structure and direction to a scenario.  Where only one point in geometry leaves open an infinite number of possible lines, only one variable in a scenario prompt is likewise not very helpful in surfacing a discrete scenario from countless possibilities.  Even two variables leave open many other considerations and thus can lead to many different scenarios, but it's still orders of magnitude better than having only one variable.  It might be desirable to underspecify a scenario anyway, e.g., if there are unknowns or to explore the permutations of a core scenario.

As to how these scenarios can be used: even building a scenario in the first place is a valuable exercise in imagineering, i.e., to think like a hacker and to begin understanding just how many unexplored possibilities exist.  At the end of the next section, we will provide critical-thinking questions to help develop a scenario from a prompt.

Once developed, these novel scenarios can help motivate discussions of current and emerging possibilities in space cybersecurity.  For instance, the scenarios could be used by multidisciplinary teams to encourage discussion across domains, as follows.  The technical team of cybersecurity researchers, aerospace engineers, and other relevant experts could be encouraged to discuss plausible attack vectors to successfully execute the attack on a real or hypothetical system.

Similarly, a policy team (considered broadly to include policy analysts as well as lawyers, ethicists, other experts) can engage in conversations around whether a particular technology or approach is within the bounds of current international and social norms.  The policy team could also formulate policy responses or solutions to novel scenarios, notably in a way that doesn't escalate tensions more than they already are.

The power of these conversations is amplified when technical and policy teams collaborate and iterate with each other.  For instance, as a technical team suggests a new feasible offensive or defensive technique, the policy team can assess the technique for its unintended consequences, including possible dual-uses and compliance with current space law and norms.  Similarly, when the policy team proposes new best practices, the technical team can develop new attack trees and potential defensive technologies that meet the best-practices proposed by the policy team.  Other possibilities include running tabletop simulation or wargaming exercises.

---

capabilities wouldn't likely impact marginalized populations who presumably wouldn't have access to those expensive services in the first place.

# Categories explained

Below is a brief description of the five categories (columns A-E) in the ICARUS matrix, followed by descriptions of each variable in the next sub-section:

### A. Threat actors or agents

The threat actor or agent is the one who causes or is responsible for launching the attack. It is worth noting that the threat actor does not need to be limited to only one of the variables or rows in the ICARUS matrix. For example, a major space-faring nation may undertake an action in *collaboration* with a minor space-faring nation or an independent corporate entity. Different actors may have different strengths, roles, and limitations in participating in a space cyberoperation, as we discuss in the next sub-section that describes each of the threat actors.

Some important cyberthreats aren't actors or agents per se, but they are *natural threats*. Examples of this include space radiation, solar flares, micrometeorites, and other such hazards in outer space that pre-date humans. Natural occurrences like solar radiation can lead to scenarios that make the operating conditions less desirable, as they can disrupt and even crash digital systems, e.g., with bit flips, just as an actual cyberattack could.

We note this because human threat actors could leverage natural threats, such as to time their cyberattack with a solar storm so that foul play isn't suspected, or to target systems that track or protect against these natural threats to expose their intended victim to those hazards.

### B. Motivations

The motivations of the threat actors need to be considered as they can point to different responses and defenses, but motivations are difficult to determine and can be complex. For example, an emerging spacefaring nation might be motivated by the actions of other states, financial interests, the personality of a charismatic dictator, or any number of motivating factors.

Motivations can also heavily influence the resources an actor is willing to commit, the perception of the action by a global audience, and the fervor the actor may act with. It is important to remember that human actions and decision-making are multi-faceted; for example, an actor that is primarily motivated by financial gain may also be motivated by professional pride, nationalism, and/or a cult of personality.

### C. Cyberattack methods

The cyberattack method refers to the specific tactic or exploit used by a threat actor. The following are three broad domains for attacks that a threat actor may use, which can be combined in a cyber campaign:

- *Cyberattacks* consist of various electronic warfare, hacking, software, hardware, and supply chain attacks. Many of these attacks will be like information technology and operational technology attacks that have been explored by the cybersecurity community over the last three decades. However, the space environment has some unique challenges that we will discuss in more detail.

- *Social attacks* involve tricking, coercing, or bribing one or more employees to do the bidding of the attacker, from gaining access to installing malware. These social attacks are extremely successful in current attacks on IT and OT system and are used to mount cyber campaigns across industries.

- *Astronautics* systems have particular and challenging operating and engineering conditions. Attacking the sensors, actuators, and algorithms unique to space systems creates a minimally explored attack-surface, at least previously to the attack, that can be exploited in novel ways.

Note that, while this report focuses more novel scenarios in the *space segment* as the more unexplored segment, cyberattacks can happen in any of the various segments of a space system. For example, a signal jamming attack can interfere with space-to-ground communications, space-to-space signals, operations and navigation of a rover, and so on.

### D. Victims or stakeholders

The victims or stakeholders are the ones who stand to lose or are otherwise harmed by a cyberattack, and knowing who the intended victims are can also help in negotiating a scenario. Like the threat actor, the impact does not need to be limited to a single victim. It is also worth noting that a threat actor who carries out one incident may be a *victim* of a different incident. For example, one space-faring nation may be the victim of a cyberattack and decide to respond in kind or escalate and launch their own attack upon another state.

### E. Space capabilities affected

We broadly define space capabilities in the following three domains: terrestrial assets that support space systems, space systems, and the terrestrial assets that are supported by space capabilities. Below are examples for each domain:

- *Terrestrial assets* to support space systems can include space launch complexes, command and control centers, communications, R&D facilities, Earth-based telescopes, end-user terminals, and more.

- *Space systems* can include crewed (e.g., shuttles, command modules, and space stations) and uncrewed (e.g., GPS satellites, probes, space telescopes, rovers) spacecraft.

- Major terrestrial capabilities that *depend* on space services can include the positioning, navigation, and timing (PNT) applications of GPS for broad civilian and defense applications, including navigation, aviation, power grid management, logistics, financial services, and more.

## Variables explained

Below is a brief description of each variable for each of the five categories above. Note that the variables are not necessarily mutually exclusive; some can overlap, which is to be expected as space capabilities can be dual-use, hacker motivations can be multi-dimensional, and so on.

### A. Threat actors or agents

Again, threat actors don't need to be independent or solitary; many actors may collaborate with, employ, or be coerced by other actors.

A1. Major spacefaring states / entities: This classification is about the more experienced and active spacefaring states or entities, which include the United States, Canada, United Kingdom, European Space Agency (ESA), France, Germany, India, Russia, China, Commonwealth of Independent States (CIS), International Telecommunication Satellite Organization (ITSO), and others. Who or what counts as a "major" spacefaring nation or entity isn't important to define here, as long as the understanding is used consistently by the users of this ICARUS matrix.

A2. Other spacefaring states / entities: This classification is based on a lower number of payloads in space as well as less experience and resources in spaceflight, as compared to the major spacefaring states/entities.

A3. Non-space-faring states / entities: This classification captures all other states and entities not in A1 or A2 above.

A4.  Insider threats:  No one likes to think that one of their own employees, colleagues, vendors, or even friends could be a threat to one's work and organization.  These insiders don't necessarily need to be malicious; they can also just be lax in their cybersecurity habits, or coerced, or pawns in a larger operation, and so on.  The insider threat is more likely to be a "lone wolf" actor, such as a disgruntled worker, than part of an organized effort.

A5.  Political terrorists:  These are violent, extremist groups motivated by specific agendas such as political change, propaganda, elections, power, sabotage, and disruption.  They could even be state-sponsored hacktivists.  Again, understanding the threat actor's motivation will be essential but those of political terrorists could be particularly diverse.

A6.  Mercenaries:  These hackers could be state-sponsored or otherwise employed privately.  As private military contractors (PMCs) continue to play a role in modern warfare, the new realities of an increasingly wired world are also creating combatants dedicated to cyberwarfare, and we can also expect cyber mercenaries to develop alongside the trend as complementary assets.[128]

A7.  Eco-terrorists:  Some extremist groups seek to either protect or harm the environment.  This group differentiates themselves from other terrorist groups with their main focus on changing the environment, but they will use violence to force a legal or policy change.  The space industry should review attacks on ground-based operations and space debris as an attack-vector.  Eco-terrorists may be inclined to protect the "space environment"—or destroy it.

A8.  Corporations:  The threat landscape that surrounds space corporations is constantly being tested for vulnerabilities from both inside and outside entities.  Private and public space corporations help run the global economy, and attacks on the commercial space industry and space-based assets are on high alert.  These vital technologies have potential cyber vulnerabilities with respect to intellectual property and proprietary data.  At least some of outer space should be considered a critical infrastructure sector.[129]  Hackers from other states continue their attempts to collect data from satellites and disrupt communications; one goal is to find vulnerabilities and exploit them during major conflicts.

A9.  Mobile service providers:  As a threat actor, mobile service providers may be much more likely than other corporations to be coerced or otherwise leveraged to conduct a cyberattack, as opposed to being motivated on their own to cause harm.  This is due to the specialized resources they can bring to bear, e.g., in a multi-step attack.  They could also be unknowingly coopted in an attack, e.g., in SIM swapping or port-out scams that aim to take over customer accounts, or spoofing cellular towers to intercept data transmissions.

A10. Launch service providers:  Large private corporations have slowly replaced NASA as launch service providers (LSP) and have changed the landscape.  Space-bound missions have many vulnerabilities because of the frequency and speed of the launches.  As with mobile service providers, LSPs have specialized resources that are attractive to hackers, so LSPs also may be more likely to be coerced or unwillingly leveraged to be involved with a cyberattack, e.g., through vendors or contractors who work remotely and have escalated privileges.

A11. Social engineering groups:  These are well-established threat actors that use persuasion and manipulation to steal data and information to use later to compromise systems as well as people.  Space assets have been targeted in recent years, upon which commerce, security, and other domains depend.  Social engineering will use tactics to get upper management to hand over credentials so that threat actors could possibly compromise sensitive data and intellectual property.  Properly trained humans as well as best practices, such as the principle of least privilege, can help defend against these types of attacks.

A12. Organized crime:  To generate illicit revenue and for other possible goals, organized crime is capable of conducting cyberoperations, such as proliferating ransomware to encrypt important data and extort the victim-organization to pay a ransom, which can be millions of dollars depending on the value of the encrypted data.  A high-value satellite or human could present a financially lucrative target for an organized crime syndicate.  Other cyberattacks may be more sophisticated and require resources and coordination that a criminal organization can provide, as opposed to a lone criminal hacker.

A13. Chaos agent:  A chaos agent, such as an anarchist, generally uses chaos and power to transform and destroy, instead of balance the societal normalcy.  The explosion in the number of annual launches in the last decade, which increases nearly every year, has created some chaos agents globally who are concerned with the growing number of satellites in low Earth orbit (LEO) and the debris that poses some sustainability in space and make some orbits unusable.

A14. Religious / apocalyptic:  Apocalyptic cults have led to unexpected and erratic group behavior, which has often been tied to religious movements in the grips of a charismatic leader.  Using as an example Heaven's Gate, a UFO/UAP religion with a charismatic leader, the cult believed they would be resurrected in a higher level of existence if they committed suicide while the Hale-Bopp comet was passing by Earth; this led to a mass suicide in a basement in San Diego in 1997.  A similar apocalyptic cult could want to target spacecrafts under the guidance of a megalomaniacal leader.

A15.  Other ideological groups:  Other extremist groups could also take violent or otherwise illegal actions in support of their goals.  Even if we are not able to currently identify or predict them, it is important to keep in mind the role of future "known unknowns."

A16.  Proxies / agents, especially unwilling:  Threat actors may want to hide behind innocent people, such as those who have been coerced or tricked into participating in a cyberoperation.  Spoofing one's IP address in the commission of a cyberattack could also implicate an innocent third-party.  This can complicate not only the attribution of a cyberattack but also responses and consequences to that attack.  Some proxies may also be willing, such as allied nation-state that is directed to hack a mutual adversary's systems.

A17.  Noncombatants, especially unwilling:  In the context of warfare, noncombatants are ordinary civilians but also military chaplains, doctors, warfighters who have been injured and taken out of action, prisoners of war, and others who are not engaged in combat.  As such, they enjoy legal protection from being targeted in a military attack.  But should they pick up arms or otherwise directly and actively participate in hostilities, they lose that protected status.  In Russia's 2022 invasion of Ukraine, Starlink was accused of participating in hostilities by providing critical satellite internet connectivity to Ukraine, effectively helping them to resist and fight Russia—and making them a legitimate target (in Russia's view, at least), even though Starlink primarily provides services to civil society.[130]

A18.  Amateur hackers / enthusiasts:  This threat actor isn't so much motivated by malice but by curiosity about whether they're able to break into a system, as a goal itself.  But given their inexperience, they could accidentally break something in that system; if the target is a space asset, that could cause severe damage, such as deorbiting.  Besides curiosity, they could also be driven by boredom, attention-seeking, or other motivations.

A19.  Artificial intelligence (AI) and machine learning (ML):  Threat actors could use advancements in AI/ML to detect new vulnerabilities to leverage against satellite system payloads or other components.  AI/ML is rapidly advancing, and prompt-writing may be how we structure the next uses of these technologies.  Research and development in these advancements will be able to analyze large sets of data, including cyber threats.

A20.  Unknown / anonymous:  While some threat actors may want credit for a cyberattack, most don't want to be discovered.  Some are known for taking special pains to hide their identities, such as the hacktivist collective Anonymous, and other threat actors remain unknown despite best efforts to identify them.  Besides not being able to attribute a cyberattack to any particular threat actor, these hidden identities can also obscure motivations, which is important for anticipating their gameplan and responding strategically—though for Anonymous in particular, their motives are generally clear and publicized.

### B. Motivations

Again, it is important to note that motivations are not mutually exclusive, and some of them may be complementary or antithetical to others.  Similarly, when multiple actors are colluding on a cyberattack, their underlying motivations may be different and diverse.

B1.  Nationalism:  Superiority in space throughout the Cold War was seen as a major source of national pride.[131]  Since then, space continues to be a point of pride for nations, for instance, when China and India both successfully sent unmanned rovers to the Moon in recent years.[132]  One motivation for attacks against space capabilities could be a sense of nationalism, for instance, sabotaging another nation to protect one's own superiority in space or to prevent that other nation from a major symbolic victory.

B2.  Dominance and influence:  Besides motivations of *political* dominance, influence, and leverage by states and others, the hacking community itself has historically had an unhealthy aggressiveness because of its inherently adversarial nature and early cultural norms.  Because of this, a potential motivation is an attacker trying to prove they are the "1337" hacker, a shorthand term for elite hacker used in early message boards.  The high-profile nature of space systems combined with their perceived strong defensive posture make them extremely appealing as a target for a hacker trying to demonstrate their dominance and gain influence in the hacking community.  More broadly, an attacker that is trying to prove themselves in various communities might choose to attack different space systems; for example, gaining control of a National Reconnaissance Office (NRO) satellite might provide strong influence in anti-American groups.

B3.  Financial and economic motivations:  At least two forms of financial or economic motivations for a potential space cyberattack are relevant here.  The first motivation is financial gain, attempting to turn their attack into a revenue stream.  A common example of a financially motivated attack is ransomware, which encrypts critical data or otherwise disables key parts of a system until a ransom is paid.  Another example is a cyber intrusion that steals intellectual property.

The second motivator is to cause financial *loss* to an adversary.  This could happen at a state or corporate level.  For instance, to cause financial loss to a corporation, a hacker may choose to target space systems that are going to cause the most financial harm, such as the James Webb Space Telescope (JWST) which costs billions of dollars. In economic terrorism, the target may be less important than the scale of the losses and ensuing societal impact, such as disruption of financial stock markets. Relatedly, cyberattacks that lead to financial losses, such as increasing the failure rate of a component, could also hamper development and mission timelines, affording time for competitors to catch up.

B4.  Fraud: Related to the previous variable, fraud is typically motivated by financial rewards, except a key difference is that perpetrators don't want their fraud to be discovered, unlike a ransomware attack that must be revealed in order to be effective. For instance, the owners of an aging satellite might sabotage their own space asset to recover an insurance payout or hide evidence of other crimes.

B5.  Employment: Many actors who participate in a cyberattack may simply be performing a job for hire, particularly government employees or contractors. The goal of such an attacker is likely to align with the official policies and statements of the sponsoring nation state, even if the attack action is never claimed by the state. Attacks such as the Stuxnet worm that disrupted Iran's nuclear program are relevant in exploring this type of threat actor.

B6.  Blackmail / coercion: Sometimes, a hacker is motivated by outside forces or influence, such as various forms of coercion, including blackmail and extortion. Consider a kidnapped loved one, or a threatening influence from a totalitarian regime, or even blackmail to expose real or perceived indiscretions. A threat actor could also seek to blackmail or extort a victim as a motive itself.

B7.  Terror: At least two strategies in terrorism with respect to space systems are relevant here. The first is attacking a space system to undermine the morale of a group of people. Space systems are obvious targets for causing harm to national pride; for instance, the Hubble Space Telescope and James Webb Space Telescope (JWST) are both symbols of national and international achievement that could be a target to cause morale loss for the United States and its allies.

Another strategy is using space assets to terrorize a group of people. Consider someone corrupting the positive goals of NASA's DART mission; the original mission explored the impact of human-made objects in redirecting an asteroid to prevent future asteroid collisions with Earth. A terrorist organization could repurpose the goals of DART to redirect an asteroid toward a target or national, political, religious, or cultural significance, either in space or on Earth.

B8.  Warfare:  Space systems are critical to a modern military for position, navigation, and timing (PNT) as well as for intelligence gathering and espionage.  Motivations here can be considered in at least three ways.  First, a state actor may aim to cripple an adversary by destroying their PNT or reconnaissance space systems.  This could cause many modern defense systems to be without critical guidance data and render many of them inert, as well as cause failures in critical civilian infrastructure like power grids, cellular networks, and civilian aviation.

Second, a hacking campaign could lead to data corruption or spoofing in critical PNT and reconnaissance systems.  Consider an adversary that wanted to hide the amassing of troops on a border from satellite reconnaissance.  If they can hack into the relevant systems and replace the images with manipulated images not containing the troops, that could have significant impact on an armed conflict.

Third, space hacking can be used to repurpose space assets for space warfare.  Consider the takeover of small satellites with propulsion, such as Starlink internet satellites.  These could be repurposed to cause a Kessler-syndrome event around LEO.

B9.  Disinformation:  Similar to disinformation for warfare purposes, threat actors may be motivated by disinformation for other motives.  Consider, for example, that a person was convinced that additional funding needs to be allocated for planetary defense, such as for a future "killer asteroid" or to guard against backward contamination; they could be motivated to spoof signals and attempt to create the appearance of an inbound threat to Earth, causing panic and new budgeting priorities.  Other motives need not involve space policy at all but merely use space cyberattacks to achieve their goals.

B10.  Espionage:  High-fidelity optical, electromagnetic, and other sensing make space systems unique in their capabilities to monitor, track, and surveil earth.  This is true of the extremely powerful and prominent NRO satellites, as well as the low-cost and prolific small satellites (CubeSats) used for scientific and industrial research.  Hacking into and obtaining control of a satellite with advanced imaging capabilities could allow a nation to conduct espionage of land-based assets that are otherwise difficult for small groups or nation states to accomplish.  While NRO satellites may go through a secure design process, the smaller CubeSats may have extremely limited vetting for cybersecurity and defense.

B11.  Sabotage:  In high-stress and rapidly evolving domains like space systems, there is always the possibility for a current or former employee to become disenchanted with an ongoing project or organization and therefore motivated to sabotage a system.  This has occurred many times already in critical infrastructure, such as when a former sewage worker used a backdoor in a system to flood City Hall, which had employed him, with sewage.  Non-insiders can equally engage in sabotage for different grievances and goals.

B12.  Extremist ideology:  Most of the other motivations mentioned can be coupled with extremist ideology to make the perpetrator even more fervent and reckless in their actions.  Consider, for example, a misanthropic terrorist, someone driven by an extremist view about the impact that humans have on the ecosystem.  If this attacker decides to hack a space launch system to prevent a launch which delivers large amounts of chemical exhaust, they may be willing to operate by whatever means necessary because of their extreme views.  Similar extremism could result for political, nationalistic, religious, or other reasons.

B13.  Cult of personality:  Technology has many big personalities with obsessive followers and detractors.  These personalities can often have immense influence over fans and sometimes make rash and wrongheaded statements without considering their statements' potential.  It is plausible that a strong personality could make a statement about the danger of a space system which, correctly or incorrectly, gets interpreted as a call to action against a space system.

B14.  Paranoia / anti-technology:  An increasingly wired world can also expect to see increasing pushback and protests over the rapid technological changes in everyday life. It's not just that change can be difficult, but there's a not-unreasonable perspective that a lack of governance and oversight has already led to real harms that will continue, such as AI bias, persistent social surveillance, and projects that distract from social and infrastructure investments, e.g., public transportation.  This fear and frustration can lead a threat actor to take drastic measures.

B15.  Boredom / trolling:  The hacking community has long been acquainted with hackers who act out of no reason other than to see if they can accomplish some tasks with an excess of free time.  This has led to prominent and seminal viruses, worms, botnets, and other attacks against computer systems.  Someone with sufficient technical skills and free time may start to ask themselves "Can I hack into a satellite?"  In penetrating a system, the bored hacker may desire to troll or play a joke on the system, such as to take over an LED on a satellite and blink an obscene message in Morse code.  The motivation here is not to be famous, not to cause damage, not to start a conflict, but simple and unadulterated curiosity, coupled with a lack of responsibility or ethics.  This type of motivation can often be repurposed and harnessed as a force for good, as seen with various reformed hackers who have transformed into productive members of industry and even defense organizations.

B16. See the world burn / chaos: Sometimes, like the bored or terrorist hackers, the chaotic hacker aims to cause loss, pain, or confusion just to "watch the world burn." This type of attacker's motivation is simply chaos, which makes them very unpredictable. Consider, for example, the bored hacker above who was able to hack into a satellite; but instead of just innocuously blinking an LED with an obscene message, the hacker decides to check whether they can deorbit one or more satellites. The implications of an unplanned deorbiting could lead to many follow-on results, including additional collisions, loss of assets, loss of essential services, such as global internet services, and other effects.

B17. Social / distributive justice: Though access to space is starting to open up, it is still very much the domain of the wealthy, which creates its own concerns, such as when they exploit governance gaps or meddle with international disputes and conflicts. Meanwhile, there's still great poverty, inequity, sexism, racism, and other injustices on Earth. Social activist groups may be motivated to target space assets to thrust their cause into the spotlight, given frustrated efforts to make better progress or on what may feel like regression. The targets could be owned by the wealthy, e.g., a vanity project in space, or organizations and assets of targeted states.

B18. Intellectual / technological demonstration: A motivating factor that can bridge many of these categories is the professional pride for the quality and effectiveness of a cyberattack, an intellectual demonstration or attention-seeking ploy, e.g., "bragging rights." This is often apparent in attacks that have high impacts, as well as attacks with government sponsors. Similarly, adversaries may want to demonstrate their capabilities, e.g., as a deterrence signal.

B19. Revenge / retaliation: Revenge is a powerful motivating factor both at an individual, corporate, and national level; it is perceived as a matter of justice for many victims. Proportionality in a retaliatory response isn't typically the primary concern for revenge-seekers, as opposed to punishing the offender and deterring other would-be offenders or restoring lost national pride. This creates a real risk of a disproportionate response, an over-reaction that could start both parties climbing the escalatory ladder toward even worse hostilities and damage.

B20. First contact, for and against: The possibility of extraterrestrial life is either exciting or frightening, or even both, depending on who you are. It would certainly be one of the greatest discoveries in human history. The extreme emotions and high-profile nature of a first contact with intelligence ET life may be enough to motivate some threat actors to spoof a first contact (when there's no evidence of ET life) or sabotage a first contact (on the off-chance that it might happen someday).

### C. Cyberattack methods

The following is a description of *general* attack classes that may be used independently or in combination in a real-world attack.  Cyberattacks typically involve a multi-step, multi-vector process that leverages multiple attack techniques.  An attacker will generally gain initial access through an external vulnerability.  The attacker will then scan and assess for additional vulnerabilities or weaknesses and pivot to an additional attack.

The details of the attack—i.e., the process of gaining access, elevating privilege, gaining persistence, and eventually achieving the attacker's desired goal—can be modeled in various ways.  For example, the [Lockheed Cyber Kill Chain](#) and [MITRE ATT&CK Framework](#) are among the most capable cyber-modeling techniques.[133,134]

C1.  Insider attack:  "Insider" refers to former and current employees, vendors, and contractors who may have had or have current access to the systems of their employer or client.  For instance, a disgruntled employee of a space company who was not considered for a promotion and feels aggrieved may be motivated to compromise digital assets, including stealing intellectual property or just causing chaos.

Important differences between an insider attack and other attacks make them a distinctly pernicious threat.  For instance, insiders may already have access to valid credentials and permissions to access critical systems, thus avoiding the trouble of developing or paying for specialized hacking skills.  Insiders also may have an understanding of system designs and weaknesses that are not public knowledge, which can contribute to more effective attacks.

C2.  Social engineering:  This is when threat actors use manipulation, coercion, and social tactics to trick users into providing information or data that will compromise them in some manner within the organization.  For instance, a phishing email is designed to trick the system administrator to divulge their credentials so that the threat actor can access sensitive data or applications.  Social engineering is often a useful technique to gain initial access into a system that can then be leveraged into additional attacks.

C3.  Ransomware:  This malware encrypts or disables one or more systems, leaving a system in a non-operative state unless the operator pays a ransom to restore the system.  Even if a ransom is paid, there are no assurances that the victim won't be milked for even more money and that the data may be deleted or not unlocked anyway.  That data is mission-critical for the day-to-day operations of many organizations; for instance, hospitals have been victims of ransomware, unable to access patient records for life-saving treatments and surgeries.  In a space system, time-critical ransom software could exist around life-support or for a deorbiting craft.

C4.  Honeypot:  This is a lure, a trove of what seems like valuable data but instead entraps the visitor.  In cyber defense, a honeypot is a sting operation; it can be a staged or sacrificial area of an enterprise IT system that contains fake data as bait.  A hacker that falls for this decoy would waste time and efforts there, while being surveilled to uncover identities, methodologies, and other such information for prosecution or other actions.  Crucially, a honeypot can be an early-warning system to detect when someone is attempting to access an information system without authorization.

As a tactic for cyber *offense*, a honeypot can similarly trick an innocent user into interacting with what looks to be a legitimate website or area of an IT system, in order to compromise the user, e.g., by installing malware on their devices.  Also known as a watering-hole attack, this is essentially a spoofed website and related tactics.  Honeytraps can also refer to entrapping a target through a romantic seduction and entanglement, such as on a dating site.

C5.  Sensor attack:  This consists of corrupting the sensor system of a spacecraft to disrupt or alter digital signals.  For example, consider a LIDAR sensor that is used to measure the distance between two spacecrafts attempting to dock.  If this sensor is compromised to read just one meter farther than what is actual, then a collision could occur during an attempted docking maneuver.  A sensor that is disabled or destroyed can also create massive disruptions to a mission.  For example, a destroyed LIDAR sensor for docking may cause the entire mission to be canceled, causing an expensive relaunch or delay.

C6.  Signals jamming:  Wireless signals enable new and exciting frontiers in space.  Whether it is deep space telemetry received from Voyager or near-space communications, including GPS, we depend on these signals for pushing the boundaries of our understanding of the universe and for modern life.  However, it is important to note that these signals travel a long way.  Even in low Earth orbit, the signal will travel 2,000 kilometers or more than 1,200 miles.  Wireless electromagnetic signals are generally modeled with a dissipation inversely squared; in other words, signals that reach the ground from space are extremely weak.  Because of this, the jamming or interruption of signals from space is easy to accomplish.

There are many products on the market that offer to illegally jam GPS signals under the guise of protecting a truck driver's privacy.  These low-cost and readily available jammers have caused problems in the past and even shut down major airports.  They often follow a constant jamming approach, meaning that they constantly send out a moderate to high power signal to accomplish their goal.  On a more optimistic note, many techniques exist to monitor and geo-locate these signals to identify and neutralize the jammer.

More stealthy and lower power jamming has also been realized in the past two decades, such as reactive jamming and adaptive jamming. In reactive jamming, an attacker listens for a signal and only jams if the signal contains some type of information; for example, an attacker may listen to traffic coming from a Starlink satellite and only jam if the signal is addressed to a specific receiver. Adaptive jamming adapts the attacker's signal to minimize the amount of energy an attacker outputs to decrease the likelihood of detection. The evolution of more intelligent jamming has increased the challenge of designing jamming resilient communications and geo-locating defenses.

C7. Signals spoofing or hijacking: Attackers can also spoof or create signals that are difficult to distinguish from legitimate signals. This type of attack typically occurs in the ground segment as more effective and easier than actually hacking a satellite. We can consider two motivating examples here. The first is spoofing of GPS signals to cause a GPS receiver to misclassify *where* it was located. This has been demonstrated in real-life to effectively trick a commercial GPS receiver, in contrast to their more secure, military-grade counterparts. While *jamming* GPS can be done cheaply and with limited technical skills, effectively and correctly spoofing GPS is much more difficult and more expensive.

The second such attack we can consider is spoofing a satellite or other device. By spoofing a satellite, an attacker could inject signals to falsify data towards whichever goal. For instance, one way to do this would be to repurpose a decommissioned satellite to pretend it's currently in use. This could cause confusion on the ground or be used to send bad sensor data back to earth.

C8. Eavesdropping or man-in-the-middle: This can be either passive or active; the former is known as eavesdropping, and the latter is known as a man-in-the-middle (MITM). In an eavesdropping attack, an attacker listens in for unencrypted or poorly encrypted communications. The attacker can then use these communications either for direct information gathering or for planning future attack pivots.

In a MITM scenario, an attacker intercepts communications and then alters or changes them. A MITM attack could either decrypt and alter packets, drop packets, reorder packets, or just send packets multiple times. Note that MITM attacks on *wireless* networks are much harder than eavesdropping attacks, as wireless or satellite signals reach an attacker's satellite dish about the same time as they reach other dishes; there is no "middle" in that scenario, other than the satellite itself.

C9. Network security: Space systems are dependent on complex computer networks. Computer network security has a robust history, including many of the specialty topics we have already mentioned.  However, depending on the network topology, we could also consider packet dropping, packet reordering, and similar network attacks that could cause poor performance.  With larger satellite constellations, new routing algorithms or protocols may need to be developed to coordinate signals, and they could lead to new vulnerabilities and cyber exploits.

C10. Hardware supply chain:  Over the last decade, state actors have demonstrated increased interest in and concern with the possibility to alter hardware in the design, manufacturing, or logistics process as groundwork for a later cyber exploit.  The design and deployment of a new board component for a flight system is an inherently complex process and depends on complex software that could be potentially corrupted at several key points.

For instance, a design could be sent to a manufacturing house, which was corrupted and replaced with a malicious design to have premature failure, backchannels, or some other fail conditions.  The board will require integrated circuits and passives components, which could be corrupted by replacing parts with incorrect parts.  The board also needs to ship between each location that it is used; during transportation, it is possible that an attacker could replace or change the component.  Corruption of the hardware is extremely difficult and expensive to discover with many proof-of-concept attacks existing in the literature.

Moreover, satellites and other spacecraft tend to sit untouched for months or even years between their flight-ready testing and actual launch.  During that extended period, a launch service provider or other actors may have many opportunities to modify or otherwise interact with the object without the knowledge or permission of others.

C11. Software supply chain:  Software used for the design, manufacturing, development, and control of space systems is likewise incredibly complex.  The design and deployment of this software involves software design tools, operating systems, compilers, libraries, and other existing code.  The integration and use of these tools depends on the security of these tools.  In the past five years, many examples of corrupted tools and libraries have been seen in the wild, which has led to compromised software.

C12. AI / ML / computer vision attacks:  The emergence of large language models (LLMs) and other advanced general-purpose artificial intelligence (AI) models have led to a fundamental shift in the approach to many tasks.  However, it has been demonstrated that attacks against machine learning (ML) can be effectively deployed; for example, an attacker can corrupt a model to misclassify an important class of data.

C13. Attack coverup:  Often an attacker tries to minimize their chances of getting discovered by covering up their actions and removing evidence of their misdeeds.  To do this, the attacker may try to delete or alter security logs or delete malware from easy-to-discover locations.  This can make post-incident forensics of an incident or hack much more difficult.

C14. Software hacking:  Modern computer software is extremely complex and involves thousands of functions with hundreds of thousands of lines of code or more.  Each function that takes user input, or input that could be derived from human input, has the possibility of performing in unexpected ways.  This can involve broad attack classes like inappropriate access control, unexpected library behavior, or corruption of computer memory, e.g., buffer overflows.

C15. Systems security:  The interconnection between software and hardware systems can result in interesting emergent behaviors.  For example, consider a set of temperature sensors distributed across the hull of a spaceship.  If the sensors malfunction and send data to the attached computer at a faster rate than the processor can handle, the processor may be overloaded and fail.

C16. Multi-phase attack / advanced persistent threat:  In an advanced persistent threat (APT), an attacker gets an initial foothold into a system.  At this point, the attacker can learn more about the system, test their defenses, and then launch attacks.  A multi-phase attack involves an intelligent threat agent using a combination of attack methodologies as outlined in this section to create a more effective attack.

C17. Cloud hacking:  As more systems rely on cloud service providers for computational and storage resources, cloud exploitation is a growing threat.  Attackers may exploit permissions errors or vulnerabilities in cloud service providers in order to compromise data relating to satellite missions and operations.  Some cloud service providers, such as Amazon Web Services and Microsoft Azure, are beginning to offer satellite ground-stations as a service, renting time on shared ground-stations for space communications.  An attacker who gains a foothold in a customer cloud tenancy through some other means, such as a misconfigured website running in the same cloud account, could potentially use this access to send command and control messages to the customer's satellite using their ground-station profile.

C18. Account compromise:  In a distributed system, accounts are used for managing permissions of who is allowed to retrieve data and control a device.  In theory, with appropriate rules for authentication and multi-factor authentication, it should be possible to securely manage accounts for users and minimize inappropriate privilege assignments.  However, users often reuse credentials, ignore multi-factor authentication, share passwords, or improperly store cryptographic keys. Account compromise is when an attacker is able to access a system through any of these means including, for example, password stuffing and two-factor authentication (2FA) bypass.

C19. Quantum computing / communications: While these technologies are still emerging, new vulnerabilities may be discovered in their use that could be exploited by hackers. Quantum computing could also be used as a method of attack, for instance, to break certain commonly used encryption schemes.

C20. Death by 1,000 cuts / long game: While not a method of cyberattack itself, this is a strategy to cause a long series of low-level damage to a system in which each incident isn't enough to raise alarm bells or trigger a response but can add up to a powerful attack.

### D. Victims or stakeholders

When considering a cyberattack, it is important to understand who may be affected, and a successful cyberattack is likely to impact a wide range of victims. This includes the obvious impact on operators of space systems, but it can also include the broad population that depends on space systems. By having an estimation of the victims, it helps cybersecurity experts to better mitigate the most dangerous cyberattacks, as well as to better understand hacker motivations.

Below is a brief description of each of the areas of potential victims and stakeholders. Again, the variables are not necessarily mutually exclusive; some can overlap, which is to be expected.

D1. Major spacefaring states / entities: See A1 above for this definition. The differences between major, other, and non-spacefaring states/entities are important since they may indicate a dependency on satellites, the number of people that are affected, and the costs suffered.

D2. Other space-faring states / entities: See A2 above for this definition. Because they have fewer space assets than major spacefaring states/entities, these other spacefaring nations may be *less* likely at risk of a space cyberattack, since they and their economies are less dependent on satellites. But the opposite may also be true: they could be targeted if they don't have robust space cybersecurity.

D3. Non-space-faring states / entities: See A3 above for this definition. With no satellites in orbit, they and their economies would be even less at risk of a space cyberattack. But keep in mind that the risk is never zero, as they could be secondary victims in a globalized world with interdependent economies and technologies. For instance, a GPS outage would affect air, maritime, and other transportation that depend on such location-based services, potentially preventing exports, imports, as well as the ability of local stores to keep their shelves restocked with food and other supplies.

D4. State-owned entities: State-owned entities (SOEs) are particularly vulnerable to attack due to their connection with the government. Many critical services are SOEs, such as public transportation systems, postal services, and petrochemical facilities. These services depend on space technology, such as the postal services' reliance on GPS signals. An attack on these services can have a direct impact on the economy and political environment. Since SOEs are controlled by the government, the security of these entities is sometimes minimal due to budget constraints, making them even more vulnerable to attack.

D5. Military and other contractors: These potential victims work to perform critical operations in the service of national security and thus are prime targets for state enemies and advanced persistent threats. Modern militaries depend on space technology, particularly satellites, for situational awareness and coordination of attacks and other activities. An attack on military and relevant contractors can have a detrimental effect on the economy, the political landscape, and our national security. Threat actors may target these organizations for reconnaissance on defense technology to gain an advantage on the battlefield. Some threat actors may also try to compromise these organizations to sway the political landscape in their favor.

D6. Scientific organizations: NASA, European Space Agency, and other organizations at the forefront of space science and research may be high-profile targets for threat actors. Possible motives include stalling such progress, e.g., if it's believed that humans should not or cannot be trusted to develop outer space responsibly, as well as simply finding a symbolic target that would bring much attention to the cyberattack. National rivalries could also motivate such attacks, e.g., to buy time for competing space programs to catch up.

D7. Corporations: Companies are very vulnerable to many cyberattacks, and there are many areas that can be targeted in their space assets. One area would be intellectual property, which can be vital to the company's fiscal responsibilities and to the shareholders. If a space corporation were hacked or suffered a ransomware attack, the outcome could be catastrophic. Worse, the attacker could then copy the stolen innovations without having to incur significant R&D costs. Large corporations need a fully staffed cybersecurity department to ward off these and other threat actors daily.

D8.   Wealthy individuals:  This demographic has more to lose and therefore makes for attractive targets.  Many famous people have been compromised by threat actors that were able to steal money or something else of value.  The wealthy are also easy targets because their personal identifiable information (PII), especially for high-profile people, may be publicly found by open-source intelligence (OSINT).  Celebrities can be notably attractive targets to boost the profile of the attack and therefore the prestige of the attacker.

D9.   General population / society:  Not just specific people and groups, but broader society can be affected by space cybersecurity attacks.  As mentioned in the introductory section of this report, space services are vital to the modern world, including for our wellbeing, ecosystems, science, transportation, communications, location-based services, emergency responders, and so on.

D10.  Indirect / secondary stakeholders:  Some victims of a space cyberattack could be collateral damage or unintended victims, such as investors in a hacked company.  These indirect or secondary stakeholders can also be leveraged in social engineering attacks due to their connection to a space company or project.  Because they do not have direct involvement with the project, they may be less aware or concerned with being socially engineered and therefore vulnerable.

D11.  Marginalized populations:  Consider a major cyberattack against space assets that are used for agriculture and food distribution.  If a cyberattack successfully disrupts food distribution, e.g., causing a lasting GPS outage, this could lead to a loss of food and a global food shortage.  In that event, the global poor could be disproportionately harmed.  As a second example, consider a persecuted population that is under the rule of a non-spacefaring nation; a cyberattack by that nation to gain unauthorized access to space feeds could give the nation a greater ability to repress as well as track the persecuted populations.

D12.  Social movements:  In recent times, social movements have been gaining more attention as major inequities and injustices persist; but even beyond physics, it seems that every action causes an opposing reaction.  Some reactions may seek to counter those high-profile social movements through cyber means, such as to target members, disrupt communications during protests, or to achieve other goals.

D13.  Cultural / religious groups:  Similar to the previous variable on social movements, cultural and religious trends also experience pushback from opposing groups, often with a different kind of fervor.  Likewise, counter-reactions to these trends may turn to illegal means, from cyberattacks to physical violence.

D14. Unions / labor representation:  While membership in labor unions has been [declining](#) for decades, the pandemic has brought work-life balance back to the forefront for many employees.[135]  Tensions between labor and management continue to be exacerbated with layoffs (including because of technological displacement, notably AI), capricious firings, sudden reversals in work-from-home policies, and other actions that are unfriendly to labor.  As unions seek to regain momentum and negotiating power, many organizations may seek to disrupt union efforts.

D15. Customers / users via their data:  While corporations can be the target of a space cyberattack, so can their customers, users, employees, vendors, and others if their data is accessed, just as with ordinary data breaches in cybersecurity.  But these demographics could be of special interest to threat actors if they are part of the space sector's ecosystems, such as the dating profiles of military officers or a space company's employees to find a particularly vulnerable target within those ranks.

D16. Individual targets:  More than a general interest in user or employee data, threat actors may be looking to target specific individuals or specific groups, such as a cohort of space tourists.  The motive could be blackmail or to steal their elevated system credentials, among any number of possibilities.  If the individual is employed at a space company, this could give unauthorized access to government-classified and mission-critical data.  If the individual is a user of, say, satellite internet, a data breach could reveal the person's location, and a high-value target during wartime would be at risk of a physical attack.

D17. Critical specialists:  These individual targets have strategic importance in the space ecosystem, from command to operations; therefore, disrupting their work can have a disproportionate impact, if they cannot be easily replaced by another person or if they are supervising a critical functionality that threat actors are looking to disrupt or deny.

D18. Critical infrastructure:  This variable spans many areas, such as water supplies, food production systems, emergency services, healthcare, transportation, and more.  Power grids are a particularly vital piece of critical infrastructure, as they supply power to all other areas.  Disruption to these services can lead to compromised data and financial losses, including loss of human lives.

D19. Internet / media / entertainment:  Internet service providers (ISP) provide connectivity to homes, commercial companies, government agencies, and defense-related organizations.  A direct attack on an ISP can impact a variety of different users.  For example, if a large space organization experienced an internet outage because their ISP was attacked, this can seriously disrupt daily operations and affect projects and missions.  Satellite internet providers, such as Starlink, can be attractive targets since they provide service to many more geographies than a typical ISP. In particular, entertainment is a large part of the modern world, such as sports and music, and those stakeholders are keen on ensuring cybersecurity events are at a minimum.

D20. AI / machine learning: Artificial intelligence and machine learning technologies are being increasingly integrated in the space industry. Some of the various purposes include satellite operations, mission planning, data analysis, and much more, and those stakeholders may become victims in cybersecurity attacks. One area that could become cybersecurity incidents are adversarial attacks that subtly but significantly manipulate input data to deceive AI/ML models. Some ways to mitigate these issues include implementing encryption and monitoring anomalous behavior. AI/ML will continue to be a part of the space industry, and it will become more defined in cybersecurity vulnerabilities.

### E. Space capabilities affected

As noted in our terminology discussion at the start of this report, we will use the term "satellite" to refer to any spacecraft and does *not* include natural satellites, such as comets, moons, and other planets. The following variables are various space capabilities that may be affected by threat actors or agencies.

E1. GPS / GNSS: Known as Global Navigation Satellite System (GNSS), these are used by other satellites and human users to determine their position on the Earth or in space. Global Positioning Systems (GPS) is a subset of GNSS and the standard in the United States; Galileo is the European Union system, GLONASS is the Russian system, BeiDou is the Chinese system, QZSS is the Japanese system, and NavIC is the Indian system. To determine the position of one's location, a minimum of three GNSS satellites must be used for triangulation and trilateration.

E2. Earth observation / remote sensing: Satellites observe the Earth through sensors and cameras, tracking such things as the weather, ocean currents, land changes (e.g., deforestation and flooding), glacier and water levels, and much more. On the environmental side, examples of these spacecraft include the Geostationary Operational Environmental Satellite (GOES) and Surface Water and Ocean Topography (SWOT), both of which are operated by NASA.

E3. Military intelligence and capabilities: Satellites are also used for security and defense purposes, such as tracking military deployments and troop movements. These reconnaissance satellites can be Earth-observing or communications and include early missile warning systems, nuclear explosion detection, optical and remote sensing, and more.

E4.    Spacecraft, robotic or crewed:  Satellites can be classified into crewed spacecraft and non-crewed spacecraft.  The International Space Station (ISS) is an example of a crewed spacecraft, while the Hubble Space Telescope and other robotic missions are examples of a non-crewed spacecraft.

E5.    Life-sustaining services:  Life-sustaining services can be divided into terrestrial and space-based services.  As mentioned in E1, GPS and satellite communications are used in daily life, including for emergency communications.  In space, for example, ransomware could potentially be used to lock or limit the ongoing life-support in a space system, which would become an extremely time-critical crisis.

E6.    Other essential services:  Space systems are needed for the synchronization of power systems, civil aviation, weather prediction (as connected to other services besides weather forecasts, e.g., for farm/land and water management), and many other civilian terrestrial services.  Attacks in this sector could degrade or even shut down those services.

E7.    Other safety of personnel and others:  This area would come up when a cyberattack affects ground systems, ground personnel, as well as other people in space, such as on the ISS.  Attacks on this sector will potentially affect the space system if the personnel are not able to perform their jobs.

E8.    Loss of sovereignty and control:  This variable is about satellites operated by another state.  For instance, satellite services (e.g., GPS or internet) could be denied during a time of need, against the wishes of the victim-state.  As another example, the state's sovereignty could be affected if satellite services enabled by other states *returned* access to communications and news that the state wanted to censor or even shut down completely, e.g., during a protest.

E9.    Earthbound services:  This variable is very broad and will encompass other variables in this category.  Examples would include remote-sensing spacecraft and weather satellites, as well as spacecraft with humans, e.g., ISS, and those used for science, e.g., Hubble Space Telescope.  In the future, there could be space services for other celestial bodies besides Earth, such as those provided by satellites that orbit the Moon or Mars.

E10.   Emergency services:  This is about satellites that are needed when there is an emergency, especially when ground-based systems are not operable, for instance, if a hurricane hits a city and knocks offline or even destroys the area's infrastructure.  Satellites could give or restore communications and news access to affected areas, as well as help to keep airports open and utilities up and running.

E11. Financial transactions: Financial institutions use GPS/GNSS to determine highly accurate timing for correct sequencing of events. A few microseconds difference can impact the profitability of a financial trade. Weak signal or no signal at all would greatly affect the financial integrity of trades and financial interactions.

E12. Mining and manufacturing: Future spacecraft will have the ability to manufacture in space and potentially mine on another planet, moon, and even asteroids. Currently these don't yet exist, but proofs-of-concept exist with robots in mining and construction here on Earth.

E13. Science capability / research: Satellites are used to improve science and research capabilities, such as Earth observations and remote sensing. But research also takes place in spacecraft, such as ISS, and can even be conducted autonomously, such as with Mars rovers. Both space and Earth telescopes are used to collect scientific data about the cosmos, such as the Hubble Space Telescope or James Webb Space Telescope.

E14. Asteroid detection systems: Spacecrafts as well as ground telescopes and radar stations on Earth play a role in the early detection of asteroids that may approach the vicinity of Earth, which can be existential threats. A false alarm via a cyber intrusion could also cause global panic.

E15. Space weather monitoring: Solar winds, space radiation, and other natural weather phenomena in space can seriously damage or entirely disable spacecraft if they're not prepared for it, as well as disrupt digital life on Earth. The satellites that monitor space weather are essentially early-detection systems to forecast inclement weather so that precautions can be taken in time.

E16. Space traffic management: Without space traffic management (STM), every space asset in orbit would be at increased risk of collision with one another or space debris; so, this is a vital capability to keep online and accurate.

E17. Space tourism / development: A cyberattack could seek to stall forward movement on space tourism, space development (e.g., off-world colonies), and other such activities. One possible motive is a belief that humans have failed to take care of Earth, our home planet, and therefore should not be permitted to wreck the space environment as well.

E18. Launch capabilities: Besides federal space launch complexes, such as Cape Canaveral and Vandenberg Space Force Base, there are over a dozen private spaceports in the US, with more on the horizon. A cyberattack that seeks to degrade launch capabilities could also target the rockets, space planes, and other launch vehicles themselves.

E19. Communications:  Especially in remote locations or during emergencies when ground-based communication infrastructure is offline, satellite connectivity can be a lifeline, such as for military command and first responders.  Severing this lifeline could mean directly endangering actual lives, in addition to lesser disruptions.  As an example of a provider, Iridium was the first company to provide satellite phone service.  Starlink operates a constellation of satellites to beam internet connectivity to Earth.  DirecTV, DISH Network, and Sirus XM Radio use satellites to transmit television and radio signals.

E20. News / social media:  Disrupting news reports and social media during critical times, such as during a natural disaster or armed conflict, not only can cause confusion and chaos, but it can also hinder an effective emergency response.  Injecting false stories into news feeds can spread disinformation and distrust for a range of goals.

......................

In the next section, we will list and briefly describe some novel scenarios as starting examples.  Again, as open-source material, they can expand discussions about space cybersecurity by enabling countless more researchers, especially without security clearances, to think through these and other scenarios—not just a few generic, vague ones—as well as technical and policy solutions to them.  By anticipating novel scenarios, we can better avoid surprises and being caught unprepared.

# 04

# Scenarios: descriptions and development

This section offers a starting list of 42 novel scenarios in space cyberattacks. Again, 4,084,000 unique combinations are possible when drawing variables (rows) from two to five categories (columns) from the ICARUS matrix. Even more scenarios can be generated with additional variables and categories of interest; further, each unique prompt can give rise to multiple scenarios, given the range of possibilities for any given variable.

The description of each scenario is brief to allow room for customization, as well as for length considerations. Each one could take several pages to explicate, and at the end of this section, we offer a starting list of questions to further develop as well as interrogate the scenarios.

Again, these scenarios aim to inform technical and policy discussions in space cybersecurity by helping to anticipate surprises, since (non-classified) discussions mainly rely on only a few generic, familiar scenarios. Without being aware of a full range of possible scenarios, technical and policy solutions may be overly broad or narrow as a result of tunnel-vision.

The primary method of organizing these scenarios is both **temporal** and **spatial**. The first axis is about *when* a particular cyberattack could be feasible, and the second axis is about the *distance* from which a cyberattack could occur relative to the Earth, that is, *where* the cyberattack might occur. The organization of these scenarios can be modified by the user as desired, as there might not be consensus on either axis. To explain the two axes:

**Temporal axis**

**A. Near-term:** threat exists now or is expected within the next 5 years
**B. Medium-term:** threat is expected to begin in 5-20 years
**C. Long-term:** threat is expected to begin in no sooner than 20 years, if ever

As always, the advance of technology is difficult to predict, and so some threats may occur sooner or later than their category would indicate. The categories, of necessity, are thus broad and approximate, not a specific prediction of the exact order in which threats will arise. Accordingly, within each chronological category, threats are *not* then sub-organized by chronology, but by location.

In the cyberattack scenarios below, it should be assumed there is no clear end-date to any particular threat in the future; the timelines given are when the threats are likely to *begin*, not end.  That is, most or all these threats are long-term in the sense that, unless we successfully devise ways to prevent them, they will continue as long as we have activities in space.

### Spatial axis

The secondary aspect of organization tracks the major locations for cyberoperations, in terms of distance from the Earth or the nature of orbit, since that may limit response-options.  The locations are divided as follows:

1. **Ground-to-space:** on or from the Earth to 100 km or 62 miles up
2. **Earthbound:** in low Earth orbit (LEO), medium Earth orbit (MEO) where GPS is, and geosynchronous orbit (GEO)
3. **Cislunar and beyond:** near or on-lunar operation, to beyond our solar system

The first category captures the fact that space systems also generally involve facilities and operations on Earth as well as in aerospace short of reaching orbit; space systems aren't only in outer space.  In space, most activities now and in the foreseeable future occur in an orbit around Earth, which is the second category.  And the third category catches everything else, from near or on-lunar operations to other planets and well beyond our solar system.

The following is a list of novel scenarios, with short descriptions in the next sub-section.

# Near-term (A)

## A.1 Ground-to-space

1. Insider threats
2. Drone retargeting
3. Data spoofing
4. Privacy
5. False flag
6. Eco-terrorists

## A.2 Earthbound

7. Gaslighting
8. Responsibility to protect

9. Symbolic targets
10. Space domain awareness

### A.3 Cislunar and beyond

11. Technosignatures
12. Biosignatures

## Medium-term (B)

### B.1 Ground-to-space

13. Ransomware
14. Space lasers
15. Horizontal launches and space planes
16. AI vulnerabilities

### B.2 Earthbound

17. Satellite constellation
18. CubeSats with thrusters
19. Space wifi
20. 3-D manufacturing
21. Kessler syndrome

### B.3 Cislunar and beyond

22. Occupy the Moon
23. Life-support systems
24. Adversarial AI
25. Heating, cooling, and power
26. Death by 1,000 cuts
27. Planetary probes and rovers
28. Technological obsolescence

## Long-term (C)

### C.1 Ground-to-space

29. Dangerous launch technologies

### C.2 Earthbound

    30. Derelict space assets
    31. Autonomous service robots
    32. Asteroid mining
    33. Asteroid redirection
    34. Fake distress call
    35. Solar-based power array
    36. Earth-crossing bodies

### C.3 Cislunar and beyond

    37. Delicate maneuvers
    38. Space radiation
    39. Space pirates
    40. O'Neill cylinders
    41. Mars and lunar settlements
    42. First-contact disruption

## Description of scenarios

The following is a brief description of each scenario above, for an idea of how such a scenario might play out in broad strokes. Again, a more developed description of each is beyond the scope of this report, though it could be valuable follow-up work. We are also underspecifying these scenarios in order to leave room for customization as desired. At the end of this section, we offer critical-thinking questions to help develop and interrogate the scenarios.

1.  ***Insider threats: the phantom menace***

    An insider is never who you think it is. For instance, it could be Jennifer, a 32-year old database manager, who becomes compromised through bribery; she's been assured compensation to meet her mother's mounting medical bills for ailing physical health. Her task is to simply inject false data to corrupt the integrity of a satellite security system—poisoning the system from within.

2.  ***Drone retargeting: hostile takeover***

    Weaponized or not, drones are highly mobile on land, sea, or air, making them threats which can be compromised for many purposes. For instance, a surveillance

drone could be hijacked to become a kinetic threat aimed at a military R&D lab, if the hacker demands are not met. With some autonomous functions, satellites are drones of sorts and likewise susceptible to being redirected at new targets, such as the ISS. Some modern space planes are designed for rendezvous and proximity operations (RPO) and support autonomous, sophisticated maneuverability akin to terrestrial drones.

3. *Data spoofing: ghost in the machine*

   Does the threat exist, or is it being faked? Spoofing isn't just about GPS and other signals, which is a known scenario, but entire data streams could be spoofed, e.g., via sensor injection, to suggest concerning or threatening behavior that does not exist, such as a military build-up on a border, or illegal online activities by a high-profile person who a malicious actor wants to sabotage. This could be done for political, economic, or other advantages. See also AI vulnerabilities.

4. *Privacy: swipe right*

   Privacy is difficult in a digital world. A military officer's private account on a dating app has been compromised, down to the height, eye color, and other physical attributes of his preferred extramarital partners. An outside group leverages him through a honeytrap scheme, introducing him to one of their agents through the dating app, and they both swipe right. Over time, his security clearances, projects, and networks are revealed through the romantic entanglement.

5. *False flag: divide and conquer*

   Confusion and deception are powerful tools, especially in a geopolitical environment where allegiances and alliances are ever more critical. For instance, states A and B have a delicate relationship for space cooperation that state C wants to damage; C does so by spoofing its IP address and other clues to implicate B in C's snooping of A's sensitive data. This disrupts the relationship and threatens cooperation on space projects. A false-flag cyberoperation can also lead to more serious conflicts, including warfare, between the unsuspecting parties.

6. *Eco-terrorists: the ghost ship*

   Suppose a large oil tanker breaks open in the Atlantic Ocean; an eco-terrorist group is able to spoof GPS signals to temporarily hide the ship's true location from the parent company, as well as to disrupt the ship's ability to radio in the accident and request assistance. The company fails to launch a timely cleanup and rescue effort, and the spill becomes more significant than it should have been. The eco-terrorists use the negative public reaction to the expanded spill to protest for significant change in oil drilling and transportation, even as the spill does billions of dollars in damage and destroys the region's biodiversity in the meantime.

7. *Gaslighting: he said, she said*

It's your word against theirs. Imagine that state A deliberately crashed one of its own satellites into another one owned by state B. Denying any responsibility, A blames a non-existent cyberattack as the cause of the "accident", which includes a full disinformation campaign to support A's story. Given that cyberattacks are usually followed by denials and murky attribution, this gaslighting—a ruse to convince a victim of something other than a fairly obvious truth—may be very difficult to disprove; as a result, A evades consequences.

8. *Responsibility to protect: new limits*

Defending a space asset could require kinetic force, such as physically stopping an adversarial satellite that threatens an RPO with one's own satellite, e.g., to rip off its solar panels. But kinetic force could fragment the adversarial satellite into hundreds of bits of space debris or more, which is already a real and growing threat to all space assets. Decision-makers will need to weigh their responsibility to protect (R2P) space assets with a responsibility to *not* create more space debris, which weighs in favor of a cyberattack over kinetic action.

9. *Symbolic targets: rebuilt in my image*

Imagine that indigenous or social movement leaders seek to incite their followers to attack symbols that the "establishment" holds dear in outer space. For example, the ISS presents not only a strategic target but a symbolic one: its presence is seen as a beacon of international cooperation and, for some potential ideological groups, as evidence of a broader worldwide government whose goals don't align with the groups. Defenders must look at the core beliefs of threat actors to determine their roadmap of attack.

10. *Space domain awareness: crying wolf*

Satellites, such as the Solar and Heliospheric Observatory (SOHO), provide warnings of incoming coronal mass ejections (CME) that range from a few days to hours, allowing protective measures to be taken on Earth, such as decreasing the electric load in power grids.  If a state-sponsored hacker creates a series of false positives or alarms in SOHO's feed, showing that CMEs will or are occurring when they in fact won't, then trust in SOHO would erode, and true warnings of actual solar storms may be ignored insofar as SOHO is relied upon at all anymore.  As a result, digital services and connectivity in an area could be disrupted; as a prelude to an invasion, for instance, the local population's ability to coordinate and repel a kinetic attack could be badly degraded.

11. *Technosignatures: METI hack*

In 1938, a radio drama instigated social panic with the possibility of an alien attack, which demonstrates that manipulating the signs—or lack thereof—of outside contact has widespread potential for chaos.  If METI's (Messaging Extraterrestrial Intelligence, an active effort growing out of SETI's mission) feeds are hacked and recordings of something sounding like language are inserted in their transcription and then leaked to the media, the fallout could include a military scramble and viral social panic worldwide.

12. *Biosignatures: wild ET chase*

Similar to the METI scenario above, a deep space mission detects biological evidence of extraterrestrial life in the ground and water samples of a distant planet.  This biosignature was fake data injected onboard the autonomous rover by malicious actors.  But the effects on Earth may be real, ranging from new plans and investments to (fruitlessly) explore the planet, to panic in the financial markets and streets.

13. *Ransomware: is this an inconvenient time?*

Data can be more valuable than money, especially if it's critical data that can't be accessed when needed.  Imagine if a ransomware attack were initiated on a rocket in *mid-launch* of a payload worth $1 billion or more.  There would be great incentives to pay a ransom of $5 million or even $50 million in bitcoin to prevent a failed mission and lost payload.  Similarly, ransom demands for spacecrafts *returning* to Earth would be very difficult to refuse, especially if they were crewed or could potentially crash-land in a heavily populated area.

14. *Space lasers: pew! pew! pew!*

With the capacity to disrupt or disable space systems from afar, space lasers can introduce a new risk where attribution and accountability are obscured.  If a hacker gains control of a powerful enough laser, it could be used in space to destroy sensors and possibly overheat solar panels on a spacecraft, for instance.  Even ground-based lasers could be used to interfere with satellite operations today.[136]  In the future, capable space-to-Earth lasers could be coopted as weapons.

15. ***Horizontal launches and space planes: the X-37***

Compared to vertical launches with rockets, space planes have many advantages, such as reaching orbit more efficiently with their gradual, horizontal ascent; they could become the status quo for launches someday.  Even if sent up by a rocket, the X-37 space plane can land like a typical airplane when returning to Earth, as previous space shuttles had.  But if it were hacked and aimed at targets on Earth, the X-37 could do far more damage than a 9/11-style attack since it could be accelerated by its own engines, in addition to the relentless pull of gravity.  Such a tactic could effectively give small states and non-state actors the capabilities of non-nuclear intercontinental missiles without incurring the massive costs in developing or purchasing them.

16. ***AI vulnerabilities: unknown unknowns***

As a rapidly evolving technology, AI is opening up new attack-vectors.  For instance, an infected AI proofreading and writing software could append a worm to shared documents, collecting log-in credentials from users.  AI is a multi-tiered threat that can range from the lightning-fast creation of disinformation to running multiple password lockout schemes.  What AI failures might look like in space are still largely unexplored.

17. ***Satellite constellation: flash crash***

Constellations are becoming more popular for their redundancy and coverage areas, e.g., effectively building a mesh network above the planet to beam down internet, phone, and television signals.  But since the satellites in a constellation are all coordinated with each other, it may be possible to take down the entire network with cyber means, including deorbiting the constellation or just a few critical satellites, e.g., to disable keystone routing points to disrupt communications.

18. ***CubeSats with thrusters: dual-use weapons***

CubeSats today don't usually have propulsion systems given their size limitations, but some are being developed with that maneuverability, e.g., with water-based thrusters.[137]  They also typically have neither the room nor associated budget for much cybersecurity if any, making those thrusters uniquely vulnerable to attack.

Such hijacked CubeSats would be dangerous if activated during their release in orbit, which may threaten the deployment vehicle, such as a rocket or space station. In the future with more capabilities, maneuverable CubeSats could become stealth anti-satellite weapon; for instance, a company may want to use the tactic to destroy one of its own failing satellites in order to recover an insurance payout.

19. *Space wifi: damn tourists*

Tourists notoriously can compromise safety, both their own and for others. In the mid-future when space tourism is popular, a few tourists go online in a space hotel to share their adventures on social media—but they click a malicious link that causes malware to be installed on the space habitat's IT systems that degrade power and life-support systems. A malicious actor, disguised as a tourist, could do the same by through malware in a USB thumb drive. Laptops on the ISS have already been infected.[138]

20. *3-D manufacturing: built to fail*

Space missions cannot launch all the replacement parts they might ever need, so 3-D or additive printers can be an invaluable resource for quickly creating parts on demand.[139] But as electronic devices, 3-D printers are also hackable. A malicious actor is able to gain access to a printer on a space station, reprogramming it to make tiny imperfections inside the parts it prints, undetectable by a visual inspection. Some of these built-to-fail components are a part of critical systems.

21. *Kessler syndrome: botnet of debris*

The previous CubeSat with thrusters could potentially be used as debris-cleaners, pushing space junk toward Earth to burn up on reentry. But if a swarm of those CubeSats were hijacked, each one in control of a piece of debris, it could become a botnet of sorts, a physical swarm that can be aimed at targets and even sparking a chain-reaction of collisions that bring us closer to the Kessler syndrome. Swarms of anything are formidable given their sheer, overwhelming numbers.

22. *Robot claim-staking: occupy the Moon*

There's a rush among competitor states to set up a research base on the South Pole of the Moon for its scientific value and possible trove of ice. Taking the view that "possession is nine-tenths of the law", especially while it's unclear how to properly stake a claim to use or occupy (but not to own) an off-planet area, one state plans to send a rover to start marking the boundaries of its planned base—but competitors sabotage the effort by jamming control and communications with the robot claim-staker, sending it wayward and potentially off a cliff.

23. *Life-support systems: poisoning the well*

A cyberattack on the food systems and shipments for, say, an asteroid mining encampment could intentionally contaminate the food with botulism. The suspected perpetrator is the company's competitor, currently in a mining rights dispute. With illness or starvation as their only options, the encampment lacks the resources to maintain the current mining efforts and withstand the sabotage; before long, they're forced to physically abandon the asteroid.

24. *Adversarial AI: not so intelligent*

AI has already been shown to be easily fooled, particularly computer-vision systems since they perceive the world much differently than humans do. For instance, researchers have already tricked such a system into mistaking a stop sign for a 45 miles-per-hour sign.[140] Likewise in space, threat actors could trick a spacecraft into misidentifying things that aren't there, such as a Martian rover into *not* seeing a cliff drop-off in front of it. Or, as an attack on Earth observation systems, adversarial patterns could be painted on rooftops of buildings to trigger a misclassification in certain image classification libraries. Beyond image recognition, other AI systems could also be gamed, such as large language models (LLMs) and generative AI used in space operations.

25. *Heating, cooling, and power: too hot to handle*

Attacks on the heating or cooling systems can be devastating to the spacecraft, as well as any crew inside. For instance, elevated temperatures can mean the engines or other critical systems are running hotter and wearing out faster. If liquid coolant is targeted to cause overheating, say, by venting it into space, the coolant can freeze and become potentially aimable bits of space debris.

26. *Death by 1,000 cuts: one bit at a time*

When you play the long game, no one sees you coming. Creating a game plan that takes years to complete can leave many open questions and no obvious roadmap to follow in response. Every attack or incident could have myriad explanations. For instance, a cyber campaign might aim to "flip a bit" during periods when solar flares are high, masking the attack as radiation damage; over time, the corrupted bits could cause a catastrophic systems failure.

27. ***Planetary probe and rovers: digital mirages***

Corrupting the data from a planetary probe to show, e.g., inaccurate atmospheric, temperature, or water readings could waste resources or even leave a mission stranded. For a Mars rover, corrupted data could falsely show that an area has significant subsurface ice and, based on that data, a mission launched to explore this site would waste billions of dollars if and when it fails to find the ice that was never there.

28. ***Technological obsolescence: failure is an option***

With the long timelines of some space missions, hardware will be at significant risk from obsolescence and lack of support, creating failures or cyber vulnerabilities. For instance, startups are behind many innovative space technologies, but if the startup fails, its customers may not be able to replace or service those technologies, including updating cyber defenses. In a low-bid situation, the quest to reduce costs could compromise the quality and durability of components.

29. ***Dangerous launch technologies: fusion, antimatter, and space elevators***

New power sources come with new risks. A powerful new fuel cell has been developed to launch rockets from the surface; for one mission, the fuel cell has been hacked to fail upon launch. The explosion destroys the ship, the platform, and the surrounding area for miles, with an effect similar to the world's biggest bomb going off on a launch pad. Or if space elevators ever come to be, their 60+ mile tether could be targeted by cyberattackers to snap off, which would do enormous damage after the super-structure collapses to the ground.

30. ***Derelict space assets: backdoors installed***

To help alleviate the space debris problem, new companies may emerge to manage, service, and even recycle derelict space assets. But potential spoofing or other means could obscure the work performed (or not performed) and cause satellite owners to be unaware of modifications done to their restored space assets, such as to eavesdrop on its communications, or to install a kill-switch that could be activated at will, or to fail at a predetermined time. Additionally, active debris removal technologies, such as with robotic claws, are inherently dual-use and could be hijacked to conduct adversarial RPO on functioning spacecraft, such as ripping off solar panels.

31. *Autonomous service robots: 1 out of 5 stars*

Taking control of future service robots can deny basic, critical support or overwhelm those systems with noise overload.  As an example, a Moon shuttle is in need of refueling, but the request-signal is spoofed and duplicated by a worm, causing every available refueler to converge on and badly damage the shuttle.  Or a request-signal could be manipulated to cause unnecessary or even harmful work to be performed on the spacecraft.

32. *Asteroid mining: here be pirates*

There's treasure in space, precious and rare-Earth metals in asteroids that could be worth billions or even much more—and space pirates are attracted to treasure chests. In one scheme, they hack a transport ship that is carrying mined resources back to Earth, redirecting it toward a remote area of the ocean where its cargo can be stolen.

33. *Asteroid redirection: rocks from gods*

NASA's DART mission had demonstrated the viability of redirecting the course of asteroids through spacecraft collision.  In the future, the orbit of an entire asteroid could be manipulated with precision by autonomous shepherds; a cyberattack could hijack those shepherds to redirect it toward Earth to cause tidal waves and other dangerous kinetic damage.

34. *Fake distress call: space swatting*

As a weapon, "swatting" is a high-intensity disruption with potentially deadly consequences.  Imagine that space guardians receive a distress call that a spacecraft has been commandeered by pirates.  Once the guardians breach the ship, typically in an aggressive manner since they presumed an active threat, they discover that the ship was actually *not* in distress; it was a fake call.  Worse than wasting rescue efforts and resources, a few personnel were killed in the initial breach.

35. *Space-based solar power array: it's getting dark*

Tapping into sustainable energy is a priority for the world now, and space-based solar power (SBSP) arrays are part of the plan: satellites that harness and beam solar power down to Earth.  But they could be hacked and reoriented to increase output power and overload the systems that it feeds, or to disrupt critical power supply.  This could threaten severe damage and loss of life in either case.

36. *Earth-crossing bodies: too close for comfort*

The possibility of a large space object, e.g., an asteroid, striking Earth is of high concern and the basis for planetary protection programs. The ability to skew information about those space objects, e.g., to obfuscate trajectory or impact possibilities, could create social panic, e.g., if the false data suggested impact on or near a major city. Mass evacuations are underway, sending the entire region into chaos which consequently overwhelms internet communications and causes widespread outages in the region.

37. *Delicate maneuvers: not another trolley problem*

The infamous [trolley problem](#) is a choice between two very bad options, typically involving a sacrifice of the few for the benefit of the many.[141] Space hackers could force such a no-win scenario: the autonomous docking operation of a large spacecraft is hacked as it slowly approaches the Lunar Gateway. It is now set to collide with the space station and damage the main docking port, crucial for receiving cargo and personnel. But if it is maneuvered, it can only be done slightly enough to avoid the docking port but likely piercing the wall of the laboratory module on the station where a scientist is working. What is the right response?

38. *Space radiation: the golden gun*

Similar to the SBSP scenario above, as solar-power technology becomes more capable, its potential as a weapon does, too. For instance, solar-light redirection through lenses has been theorized as a potential propulsion for deep space travel. Those same lenses could concentrate and direct solar radiation at critical space assets or even terrestrial ones, leading to overheating, electronic malfunction, and physical damage.

39. *Space pirates: private military contractors*

Pirates and mercenaries are age-old challenges that consistently adapt to new technologies and frontiers, and this means competency in space cyberoperations. In the distant future with bases and settlements on other planets, pirates could sever communications and misdirect supply transports for their own gain. Further, unscrupulous space barons aiming to establish their own autonomous empires, similar to seasteading, could hire mercenaries to defend their claims in space, as well as attack and disrupt the operations of their competitors, by both physical and cyber means.

40. ***O'Neill cylinders: more technology, more problems***

Known as an O'Neill cylinder, a rotating, cylindrical habitat that is designed to support human settlements in outer space would clearly be a complex technological ecosystem. It would integrate crucial systems for orientation, atmospheric control, and artificial gravity—each posing novel and critical risks if sabotaged, since human life would be at stake. Other space habitats could similarly be attractive targets.

41. ***Mars and lunar settlements: Internet of Things***

Off-world habitats may look like smart homes as we depend more on automation, especially when compensating for fewer people available to run and maintain the complex environment. Consider a distributed denial-of-service attack that disables, say, electronic door locks across a space settlement, similar to how DDoS attacks can crash websites as well as physical systems, such as Internet of Things or connected devices. Some settlers are locked out (which could be fatal given the freezing nightfall), others are locked in, and movement around the settlement has been halted. Attacks on other systems, such as for life support, could cause harm more directly.

42. ***First-contact disruption: you go in pieces***

Perhaps someday we will discover that humanity is not alone in the universe, via contact with another advanced civilization. But hackers could exploit communication channels between Earth and this extraterrestrial civilization to propagate false information and create misunderstandings that provoke interstellar hostilities. Adversaries might also compromise the security of first contact systems, leading to distrust—or compromise the signals we send into space, hiding a threatening message inside.

## Questions to develop the scenarios

To be more useful, each scenario above can be more fully developed since those details may matter. As an overview of the exercise, thinking critically about space cyberattack scenarios means to answer questions about basic elements: who, what, why, when, where, and how. In section 3, we explained that the ICARUS matrix is aimed at these key dimensions. Again, the five columns A-E represent the major elements of a space cyberattack scenario:

A. Threat actors or agents (**who** is perpetrating the cyberattack?)
B. Motivations (**why** are they launching a cyberattack?)
C. Cyberattack methods (**how** would the attacker penetrate a system?)

D.  Victims or stakeholders (also related to the **who** question)
E.  Space capabilities affected (**what** is the damage or effect intended by the attacker?)

It was noted previously that **where** the cyberattack takes place and **when** are determined by first establishing the other elements; they're not so much variables to select but *dependencies* of the other elements of a scenario.

With some idea of a scenario in mind, we can begin to develop and then interrogate the scenario with key critical-thinking questions, such as below. This is only a starting list of general questions, not a comprehensive one. Other questions may depend on the scenario as well as answers to previous questions. A more complete discussion on how to imagine and develop scenarios is beyond the scope of this report, but a prompt that captures the main elements already provides a running start and may even be enough for some purposes.

### Who?

- Could there be other plausible threat actors for this particular attack?
- Could an uninvolved third-party be implicated in the attack, and how?
- Are nation-states implicated as a special consideration for escalation?
- Are there ways to attribute or confirm a cyberattack has occurred, and that it was committed by a particular actor?
- Could "gaslighting" be possible, where threat actors deny their attack, and how?
- Who exactly would be the directly and indirectly affected parties by the attack?

### Why?

- Could there be other plausible motivations for the attack?
- What clues does the motivation give about the threat actor's attack, such as: how far are they willing to go, what/who are they after, how might they respond to a defender's responses, etc.?

### How?

- How would such a cyberattack be performed?
- What specific methods would be employed?
- What resources would be required to carry out the attack?

### What?

- What space assets would be involved?
- What space capabilities would be affected?

- What are the possible costs, such as: physical harm, financial harm, psychological harm, relational harm (e.g., family, friends), communal harm, societal and cultural harm, environmental (incl. animals), and future interests?
- How bad would the impact be, in terms of: quality, quantity, and probability?

**Where?**

- Which segments of the space ecosystem would be involved?
- Where would such a cyberattack take place, such as: on Earth, or in orbit in LEO to GEO, or in a higher orbit in cislunar out to LaGrange (including on the Moon itself), or beyond to interplanetary or even interstellar space?
- Does the distance from Earth limit what either the attacker or defender can do, including as responses?

**When?**

- Is the cyberattack plausible in the near, mid, or distant future?
- Could a version of the cyberattack occur earlier, and how would the scenario change?

**How to respond?**

- Since different stakeholders have access to different resources, what resources are available in responding to the attack?
- What technical or design solutions can help prevent or mitigate such an attack, or to mitigate its effects?
- What policies should be implemented to help deter future, repeated attacks?
- What kinds of sanctions or responses should be considered?
- Would a considered response be proportional to the cyberattack?
- What might be the responses to our responses?
- If de-escalation is a goal, where are the off-ramps given a particular response?
- What can be done about attempts at gaslighting or otherwise falsifying attribution, and associated denial of responsibility?

Further, to truly think critically about the scenarios, a diversity of perspectives is needed to guard against groupthink and cognitive biases, especially since threat actors are already vastly diverse groups, ranging from the apathetic to the zealot. Different disciplines are trained to address problems from different directions and methodologies. Thus, critical thinking through interdisciplinarity is a practical approach to scenario analysis and problem-solving, as it brings the distinctive expertise and power of many disciplines to bear on the task.

For instance, social scientists, such as from science and technology studies (STS), provide useful tools to uncover and examine ethnic, gender, disability, indigenous, and other issues related to technical systems.  Psychologists and other behavioral scientists can offer insights into the social engineering aspects of the scenarios.  Philosophers can bring deep analytic and conceptual skills to help frame, extend, refine, organize, and critically press on relevant issues.  Science-fiction writers and futurists are essential for imagining the unknown, often more creatively than academics can.  And of course, engineers and technologists are the architects of the systems targeted by cyberattacks; therefore, they are invaluable for assessing the mechanics of an attack and working toward a solution.

## 05

# Conclusion

On the surface, space security might not seem to be much of a problem, as there hasn't been a major incident, such as kinetic attacks on adversarial satellites. While some space cyberattacks are making the news, general awareness of them is still low for many possible reasons. Besides perhaps being too technical or obscure for most people, space cyberattacks aren't nearly as dramatic as a kinetic attack would be, and they don't typically affect people at scale.

The notable exception, of course, were those populations affected by Russia's alleged hacking of Viasat's satellite internet services at that start of its 2022 invasion of Ukraine, which also caused outages for tens of thousands of users across Europe.[142]  That incident is likely a harbinger of things to come, setting a very dangerous precedent in space cybersecurity.

Because large organizations are targeted by cyberattacks daily, it can be a numbers game and therefore only a matter of time until a serious breach in space cybersecurity disrupts life on Earth.  For instance, widespread GPS outages can affect much more than positioning and navigation—which is essential for aerospace and maritime navigation, as well as everyday transportation of goods—but also *precision timing*, as a critical need for financial transactions and modern communications.[143]

But to anticipate those future cyberattacks on space systems, we need to have more than some vague notion of GPS or satellite hacking in mind, as serious as those two scenarios may be.  A failure to imagine more possibilities can promote tunnel-vision at the expense of countless other scenarios.  This would be very bad for cyber defenders, as attackers can already be counted on to be exceptionally inventive and resourceful.

To help close that gap and prime the imagination-pump for cyber defenders, this report offers a roadmap (or star map) to many more possibilities: to more threat actors, more motivations, more cyberattack methods, more victims, and more space capabilities affected.  With our scenario-prompt generator, the ICARUS matrix, over 4 million unique prompts are possible, and each prompt can also lead to multiple scenarios.  Users can also add their own variables of interest to generate even more scenario possibilities, and other variables will likely emerge over time.

The ICARUS matrix also serves as a taxonomy of sorts that's more accessible to the broader audience intended here.  This is in contrast to existing taxonomies that are either too technical and comprehensive or too simplistic and limited for our purposes here.  The value of a taxonomy lies in its ability to provide a conceptual framework or organizing logic around a domain, such as the Linnaean taxonomy for biological classification.  With a framework in place, researchers can methodically work through a domain, as well as to identify or track certain elements for study, map their relationships to other elements, discover patterns in the domain, and so on.

By making a much larger set of space-cyberattack scenarios available to a wider range of researchers beyond those with security clearances, this work can help to better avoid surprises with more experts studying the problem and working on proactive fixes.  Policy researchers also have a role here in helping to fill in legal gaps that can be a contributing factor to space conflicts, including to develop policy responses to novel scenarios that won't escalate tensions more than they already are.  Even the more speculative scenarios can help to think through proposed legislations and treaties for any unintended consequences and gaps.

*The diverse threat actors, motivations, methods, and so forth that can be involved in future cyberattacks require diverse experts to understand them.*

As an introductory guide to "imagineering" more scenarios, the ICARUS matrix and this report are not so detailed as to be a *how-to* manual for malicious or black-hat hackers who, it must be assumed, are already thinking creatively about their next exploits.  Rather, this discussion is meant to raise general awareness and help researchers contribute to the *defense* of space systems, as part of the natural co-evolution of hunter and prey in cybersecurity.

We would hope that rational minds would prevail and understand that kinetic conflicts in space are simply unsustainable, at least as a critical environmental risk.  If open war breaks out in orbit, then all bets are off, including potentially the ability for anyone to safely access space again, given the incredible amounts of orbital debris that could fall out on the way to realizing the Kessler syndrome.

For that and other reasons explained in this report, cyberattacks can be expected to be the prevailing form of hostilities and conflict in outer space; at the least, they demand urgent attention.  Given the essential role that space systems and services play in our modern world, even if that's invisible to many or most people, space cyberattacks can have an outsized impact here on the ground.

And if the value of space systems is invisible to many or most people, then the problem of space cybersecurity may very well be, too. The types of cyberattacks on space systems may someday, before too long, become as diverse as cyberattacks on information systems here on Earth. Indeed, there doesn't seem to be any reason to think they will be limited to just a handful of types, especially given the wide attack-surface possible for complex space systems and the different types of space missions.

The diverse threat actors, motivations, methods, and so forth that can be involved in future cyberattacks require diverse experts to understand them. Not just technologists who are at the middle of this puzzle and are clearly essential, but also policymakers, lawyers, social scientists, philosophers, creatives, other humanists, and so on, such as possible victims and stakeholders, are likewise indispensable in working toward a deeper, holistic understanding.

With a wider range of novel and surprising scenarios, this report seeks to help attract and inspire more researchers, technologists, policy professionals, and the broader public to engage with the problem of space cyberattacks. We are *all* stakeholders and potential victims here. Judging from the relentless and shape-shifting cyberthreats that persist here on Earth, we will need all the help we can get to secure our space systems, the next frontier for cybersecurity.

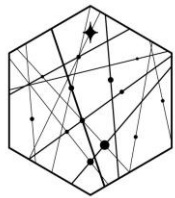

# Appendix A: the ICARUS matrix

**ICARUS matrix:** generating novel scenarios in outer space cybersecurity

*Instructions: Pick a variable from two or more columns to construct a scenario. This is not a comprehensive list but only a starter kit.*

| | A: Threat actors | B: Motivations | C: Cyberattack methods | D: Victims / stakeholders | E: Space capabilities affected |
|---|---|---|---|---|---|
| 1 | Major space-faring states | Nationalism | Insider attack | Major space-faring states | GPS / GNSS |
| 2 | Other space-faring states | Dominance / influence | Social engineering | Other space-faring states | Earth observation / remote sensing |
| 3 | Non-space-faring states | Financial / economic | Ransomware | Non-space-faring states | Military intelligence and capabilities |
| 4 | Insider threats | Fraud | Honeypot | State-owned entities | Spacecraft, robotic or crewed |
| 5 | Political terrorists | Employment | Sensor attack | Military and other contractors | Life-sustaining services |
| 6 | Mercenaries | Blackmail / coercion | Signals jamming | Scientific organizations | Other essential services |
| 7 | Eco-terrorists | Terror | Signals spoofing or hijacking | Corporations | Other safety of personnel / others |
| 8 | Corporations | Warfare | Eavesdrop / man-in-the-middle | Wealthy individuals | Loss of sovereignty / control |
| 9 | Mobile service providers | Disinformation | Network security | General population / society | Earthbound services |
| 10 | Launch service providers | Espionage | Supply chain, hardware | Indirect / secondary stakeholders | Emergency services |
| 11 | Social engineering groups | Sabotage | Supply chain, software | Marginalized populations | Financial transactions |
| 12 | Organized crime | Extremist ideology | AI / ML / computer vision | Social movements | Mining or manufacturing |
| 13 | Chaos agents | Cult of personality | Attack coverup | Cultural / religious groups | Scientific capability / research |
| 14 | Religious / apocalyptic | Paranoia / anti-technology | Software hacking | Unions / labor reps | Asteroid detection systems |
| 15 | Other ideological groups | Boredom / trolling | Systems security | Customers / users via their data | Space weather monitoring |
| 16 | Proxies / agents, esp. unwilling | See world burn / chaos | Multi-phase attack / APT | Individual targets | Space traffic management |
| 17 | Noncombatants, esp. unwilling | Social / distributive justice | Cloud hacking | Critical specialists | Space tourism |
| 18 | Amateur hackers / enthusiasts | Intellectual / tech demo | Account compromise | Critical infrastructure | Launch capabilities |
| 19 | AI / machine learning | Revenge / retaliation | Quantum computing / comms | Internet / media / entertainment | Communications |
| 20 | Unknown / anonymous | First contact, for and against | Death by 1,000 cuts / long game | AI / machine learning | News / social media |

# Endnotes

## Contact information

Prof. Patrick Lin
Director, Ethics + Emerging Sciences Group
California Polytechnic State University
Philosophy Department
1 Grand Avenue
San Luis Obispo, CA 93407
palin@calpoly.edu

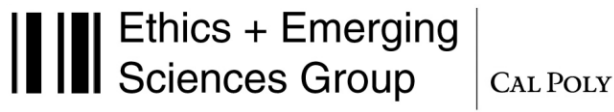